\documentclass[aps,longbibliography,notitlepage,floatfix,11pt,tightenlines,nofootinbib]{revtex4-1}

\usepackage{bm,amsmath,amssymb,graphicx,stmaryrd,subfigure}
\usepackage[abs]{overpic}
\newcommand{\uvc}[1]{\bm{\mathrm{\hat #1}}} 

\usepackage{tensor}

\newcommand{\bX}{{\bf X}}
\newcommand{\bR}{{\bf R}}
\newcommand{\bT}{{\bf T}}
\newcommand{\bF}{{\bf F}}
\newcommand{\bI}{{\bf I}}
\newcommand{\bP}{{\bf P}}
\newcommand{\bS}{{\bf S}}
\newcommand{\bC}{{\bf C}}
\newcommand{\bB}{{\bf B}}

\newcommand{\bb}{{\bf b}}
\newcommand{\bu}{{\bf u}}

\newcommand{\BH}{{\scriptstyle B_H}}
\newcommand{\BK}{{\scriptstyle B_K}}
\newcommand{\B}{{\scriptstyle B}}

\begin{document}

\title{Some observations on variational elasticity and its application to plates and membranes}
\author{J. A. Hanna}
\email{hannaj@vt.edu}
\affiliation{
Engineering Mechanics Graduate Program, \\
Virginia Polytechnic Institute and State University, Blacksburg, VA 24061, U.S.A.}

\date{\today}

\begin{abstract}
	Energies and equilibrium equations for thin elastic plates are discussed, with emphasis on several issues pertinent to recent approaches in soft condensed matter.
	Consequences of choice of basis, choice of invariant strain measures, and of approximating material energies as purely geometric in nature, are detailed.
	Ambiguities in the definitions of energies based on small-strain expansions, and in a typical informal process of dimensional reduction, are noted.
	A simple example serves to demonstrate that a commonly used bending energy has undesirable features, and it is suggested that a new theory based on Biot strains be developed.
	A compact form of the variation of a plate energy is presented.
	Throughout, the divergence form of equations is emphasized.
	An appendix relates the naive approach adopted in the main text with standard quantities in continuum mechanics.	
\end{abstract}

\maketitle

\raggedbottom

\section{Introduction}

This paper is intended to both clarify and raise issues related to approaches to hyperelasticity that are increasingly being adopted within the 
 soft condensed matter physics community, primarily for their application to the incompatible elasticity of plate and shell structures.
While aimed at this particular community, it might also stimulate some discussion with elasticians who employ approaches aligned with modern continuum mechanics, a field whose notation and attitude has diverged from physics since the mid-20th century.
The scope is limited to a comparison of a few choices of small-strain elastic energies, and the application of one of these to derive equations of equilibrium and free boundary conditions for plates. 
The latter derivation is presented in a compact vector form that I believe is not to be found elsewhere, and which I hope some will find useful.
The major points made, and their locations in the paper, are briefly listed at the end of this lengthy introduction.

I will approach these topics in a purposefully naive way, without invoking the deformation gradient or established stress tensors of continuum mechanics, and only occasionally referring to a few such strain tensors and their invariants\footnote{Here invariance will be understood to mean that under general coordinate transformations, rather than the more restricted Cartesian sense found in some continuum mechanics texts.} to indicate a correspondence with quantities at hand.
An exception to this can be found in Appendix \ref{continuum}, which translates between the physics and mechanics languages.
 The intent is to connect with the recent soft matter literature and its notation, so that the ideas expressed here can be adopted there.
This is not to say that I favor such approaches over those developed in continuum mechanics over the last three quarters of a century; 
  I think there are good arguments for both the physicists' and the mechanicians' perspectives.\footnote{I claim to be neither physicist nor mechanician but merely, to borrow Truesdell's usage, an idiot \cite{Truesdell84}.}
But I also think it should be possible to obtain results consistent with modern continuum mechanics by employing the physicists' language and approach, already well entrenched in the context of other field theories such as electromagnetism and gravity.

Proceeding in this manner, certain aspects of the derivation and presentation may seem either obvious or ridiculous to some mechanicians.  Many of the problems faced by the physicist working from his own conception of first principles, 
some such problems surely being artifacts of his approach, 
were already thought through and resolved, or avoided, in early foundational work in mechanics in the 1950s \cite{Truesdell52, GreenZerna50, GreenZerna92, DoyleEricksen56}.
The physicist 
would no doubt find the style and notation of some of these works far more accessible than the more recent continuum mechanics literature, but 
all too often takes as authoritative reference the 
 elasticity volume of the Landau-Lifshitz Course of Theoretical Physics  \cite{LandauLifshitz86}.  Though first published in 1959, it
 unfortunately does not incorporate the relevant material from the preceding decade, and treats most of elasticity as a linear field theory.

In keeping with a primary audience in physics, I will assume basic knowledge of differential geometric symbols and index gymnastics in the early 20th century style of ``absolute differential calculus'', but no background in continuum mechanics.
I will follow similar styles of notation as found in the older elasticity literature \cite{GreenZerna92, Koiter66}, some later mechanics literature derived therefrom \cite{Steigmann90, HilgersPipkin92qjmam, HilgersPipkin92qam, HilgersPipkin96}, and recent physics literature \cite{Efrati09jmps, Dias11, GengSelinger12}, but will introduce a bit of new notation in an attempt to avoid ambiguity in certain common expressions. 
Even within this stylistic slice of the literature, it is impossible to be consistent with everybody, as things like capitalization or barring of quantities are used in opposing ways by different authors.

Continuum mechanical approaches often consider very general constitutive relations or stored energy functions.
While this generality is powerful, and avoids certain problems encountered in physics approaches, it can also hide some important complexity, particularly if one wants to connect with a reduced-dimensional theory of a plate or rod.
   The alternate approach I borrow from physics is to construct a Lagrangian as an expansion in elastic strains, expressed as a functional of a position (embedding) vector and its derivatives with respect to material markers.  To these must be added other fields describing a reference configuration.  Compatibility equations are avoided because of the explicit use of position.
Stresses are derived quantities arising naturally inside a divergence.\footnote{My impression is that, for many continuum mechanicians, the stress is seen as a fundamental object, because the description of many real materials is not immediately derivable from an action principle.  Then when variational principles are invoked, perhaps for their utility in a numerical scheme, the stress may appear explicitly inside the Lagrangian.}
The first derivative of position is a tangent vector to the body, a viewpoint that seems most natural when considering low-dimensional bodies such as surfaces, but need not be so restricted.  We can think of the components of this vector as carrying one material and one spatial index.
A related fundamental quantity in continuum mechanics is the deformation gradient, a two-point tensor with one foot in a reference configuration and one in the present configuration. 
Commonly used index-free notations, matrix representations, or treatments based on Cartesians rather than convected coordinates,
 often obscure the special nature of this object that does not live in a single space.

The supposed justification for a small-strain expansion, other than physicists' seeming compulsion to write everything that way, is that the desired end is a reduced energy for a plate or shell structure in which both mid-surface stretching and the product of curvature and thickness are small. 
The energy of a two-dimensional plate derived from three-dimensional bulk elasticity will contain bending terms associated with both extrinsic and intrinsic curvatures, as well as stretching terms describing in-plane deformations of the mid-surface.  
However, it will become clear in subsequent discussion that the definition of a ``quadratic'' energy is ambiguous and potentially misleading, as one is typically choosing among common strain measures that do not have equivalent dependencies on variables like displacement derivatives or stretch.
Higher-order differences in bulk elasticity are not only important for understanding theories that extend to moderate strains, but actually
have significant lower-order effects on derived bending energies. 
Surprisingly, the most common choices of bulk strain energies do not correspond to the simplest phenomenological direct theories of lower-dimensional objects. 
These points are illustrated in a simple example, that of extensible \emph{elastica}, sketched in Section \ref{extensibleelastica} and fleshed out in another publication \cite{WoodHanna19}.

One way to view elasticity is as a theory with multiple metrics.  But a metric has two distinct uses.
One is to construct geometric quantities like the integration measure and covariant derivative, and the other is to construct invariants, for example by applying the inverse metric to derivatives of position.  As will be discussed, these are effectively choices about coordinate bases and constitutive laws.  Theories will be displayed in which the metrics of the reference or present configuration are 
used for both purposes (Sections \ref{elasticityreference} and \ref{elasticitypresent}), and in a hybrid approach in which the reference metric is used only for invariants and the present metric for everything else (Section \ref{elasticityhybrid}).  
That we are free to make such choices was recognized in early work on nonlinear elasticity \cite{GreenZerna50, GreenZerna92, GreenShield50, John65}, 
and has been recognized implicitly or explicitly in the literature on general-relativistic elasticity \cite{CarterQuintana72} and fluid membranes \cite{Jenkins77siam, Steigmann99arma}.
One finds divergence expressions employing the covariant derivatives of reference and present configurations in studies of classical \cite{GreenZerna92, Leonard61, Koiter66, Niordson85, Pietraszkiewicz83} 
  and general-relativistic \cite{KarloviniSamuelsson03} elasticity. 
The present configuration is more common, likely because of its clearer connection with boundary loadings. 
However, the geometry of the reference configuration has been favored in recent physics literature such as \cite{Efrati09jmps} and other work it has influenced.
A hybrid approach is most common in continuum mechanics. 
John's paper \cite{John65} and Green and Zerna's book \cite{GreenZerna92} suggest that either metric could be used to construct invariants, but both only use the reference metric in practice.  
In an earlier paper, Green and Zerna \cite{GreenZerna50} use both metrics to form mixed strain components, but use the reference metric to explicitly construct a Taylor expansion for an energy.  In later papers with Naghdi and others \cite{Green65, Green68, Green74-1, Green74-2}, Green has dropped any mention of the possibility of using the present metric for invariants.

What does this actually mean?  
We have a reference (rest, target) metric corresponding to a basis of tangent vectors to the reference configuration, and a present (current, deformed, actual, realized) metric corresponding to a basis of tangent vectors to the present configuration. 
Both bases are coordinate bases corresponding to the \emph{same} material coordinates, one basis being convected into the other by the deformation.  
With regard to geometry, it seems sensible to use each metric \emph{as a metric} in its own configuration; outside of its configuration, the metric is just some tensor. 
With regard to invariants, there seem to be reasons to favor the reference configuration if it is indeed a ``rest'' configuration, but really this is just a constitutive assumption.
While I reserve explicit definitions and discussion for Section \ref{elasticitynotation}, for now I point out that objects
such as the metric differences $\epsilon_{ij} = \frac{1}{2}\left( g_{ij} - \bar{g}_{ij} \right)$, formed by taking derivatives of present and reference positions, can be covariant components of either Green or Almansi strain tensors and can be acted on by either (inverse) metric to form invariants of these respective tensors.
Thus what a physicist might call a choice of metric would be more properly called a constitutive law for a hypothetical material.  The choice of basis to accompany these components is a choice of tensors --- component functions do not uniquely identify a tensor in elasticity, so a common physics shorthand may lead to serious confusion.
The reference basis is usually favored in continuum mechanics, and typically researchers have a better intuition for the meaning of Green strain and its invariants than for other options.
However, in proceeding from bulk elasticity to the inherited bending behavior of thin structures, there are clear advantages to using the present basis, because one gets to work with the proper geometric invariants of surface curvature tensors and the like.
This, along perhaps with the general lack of reference bases in other fields of physics, and the enticing potential for parallels with geometric theories in so-called ``fundamental'' physics, has tended to favor the present basis in the work of many physicists.
However, as discussed below, one needs to be careful making parallels, as elasticity is not a geometric theory in the sense intended in these other areas.
 
Using the present metric for invariants is rare in continuum mechanics.  An early work on rods by Hay \cite{Hay42} uses the present metric for all purposes, although he does not ever discuss invariants of strain, only of stress.
Volterra \cite{Volterra56} uses the present metric 
and constructs a rod energy from a strain linear in covariant derivatives of displacement. 
  Ericksen and his collaborators Rivlin \cite{EricksenRivlin54} and Doyle \cite{DoyleEricksen56} take it for granted that one should use the reference metric, as the strain measure they are applying its inverse to is simply the present metric, and the other option would just generate the identity.
However, Doyle and Ericksen \cite{DoyleEricksen56} also allow for the possibility of constructing energy expansions using quantities akin to those obtained using the present metric. 
In fact, algebraic relationships hold between the three invariants of Green and Almansi tensors \cite{DoyleEricksen56, GreenZerna92}.

Maugin's opinion \cite{Maugin95} was that the deformed metric is the right metric to use ``on the material manifold'', and Toupin \cite{Toupin57} appears to be saying the same thing, but both authors are concerned with constructing theories consistent with relativistic physics.
Rayner's \cite{Rayner63} use of a reference metric is atypical in this area.
 Oldroyd \cite{Oldroyd70} uses the convected metric to construct invariants 
for general-relativistic rheological equations of state, and Maugin \cite{Maugin71} indicates 
 use of the inverse deformation gradient to construct an energy in general-relativistic magnetoelasticity.\footnote{If you can figure out what Grot and Eringen \cite{GrotEringen66-1, GrotEringen66-2} are doing with regard to invariants in their work on special-relativistic continuum mechanics, 
please contact me.}  
This is also implicit in the work of Hernandez \cite{Hernandez70}, although he does not discuss invariants, and Beig and Schmidt \cite{BeigSchmidt03}, and explicit in papers by Kijowski and Magli \cite{KijowskiMagli92, KijowskiMagli97},
who confusingly refer to the inverse deformation gradient as the relativistic deformation gradient.
Carter and Quintana \cite{CarterQuintana72} use the present metric, which they consider to be a projection tensor of sorts, in their general treatment of general relasticity, but then use the reference metric in an isotropic Hookean approximation, and clearly recognize that one is free to choose when constructing expansions.  Karlovini and Samuelsson \cite{KarloviniSamuelsson03} follow Carter and Quintana and use the present metric, while noting freedom of choice and low-order equivalence in an appendix.

In the context of bending of surfaces, Peterson \cite{Peterson09} uses the reference metric as the metric and the integration measure, but constructs a phenomenological bending energy by using both inverse present metric and inverse reference metric to construct a difference between two curvatures.  He indicates that this choice is purely for illustrative purposes, and indeed if one were to apply the same approach to the metric tensors, the bulk strains would vanish trivially.
Similar approaches to bending can be found elsewhere \cite{NaghdiNordgren63, Flugge72}.  At first glance, the resulting object doesn't seem to correspond to an invariant of a single tensor.
Similarly, Stumpf and Makowski \cite{StumpfMakowski86} use the reference metric on the bulk and mid-surface strains in shells, while using the differences in mean and Gaussian curvature with one value scaled by the ratio of integration measures. 
Rosso and Virga \cite{RossoVirga99}, without mention of reference configurations, discuss general energies for lipid membranes that depend on curvature invariants, 
 while Maleki and co-workers \cite{Maleki13} discuss general energies for lipid membranes that may depend on curvature invariants for both present and referential surfaces.
In contrast, despite their notation (following \cite{Armon10})
and some of the statements in the discussion around their equation (2.1), Pezzulla and co-workers \cite{Pezzulla17}
 construct their shell bending energy using the reference metric and derivatives of present and reference positions, and thus are not employing the present curvature tensor except in those cases, common in their work, where the reference and present metrics of their mid-surface coincide.

Different choices of linear constitutive relations between nonlinear stresses and strains lead to qualitatively different results in bulk elasticity at large strains \cite{Batra98, Batra01}.
It will be shown in Section \ref{extensibleelastica} and a separate publication \cite{WoodHanna19} that constitutive choices have even more marked qualitative effects on derived bending elasticity.  In particular, the Green form of the bulk energy favored in some recent works \cite{Efrati09jmps, GengSelinger12, Pezzulla17} appears to generate bending energies in qualitative contrast with intuitively sensible definitions even at moderate strains.

This paper is restricted to the relatively simple case of isotropic compatible or incompatible elasticity as considered by \cite{Efrati09jmps} and related works, where it is assumed that certain quantities, such as metrics, can still be defined even if no reference configuration exists in $\mathbb{E}^3$.
  Also notable are recent studies of nematic glass or nematic elastomer sheets that involve coupling between the deformation gradient and a director field
 \cite{Modes11, NguyenSelinger17}.
These are examples of a class of complicated systems including materials with defects and inclusions, and processes of possibly anisotropic growth or swelling and subsequent elastic deformation.
Other tools have been developed, such as the ``elastic metric'' of Nardinocchi and co-workers \cite{Nardinocchi13}, or the more complicated mathematical machinery for calculus on manifolds \cite{StumpfHoppe97} employed by Yavari and co-workers \cite{Yavari06, OzakinYavari10}, who have also made an effort to relate such formalisms with existing approaches and notations.
The paper \cite{Yavari06} also discusses the distinction between the general covariance principles taken for granted in physics and the concept of objectivity under special Euclidean transformations that has taken prominence in some approaches to continuum mechanics. 
Other concepts I will not employ in the present work include the two-point ``shifter'' tensor of early continuum mechanics \cite{EricksenTruesdell57} 
for translation between two different coordinate systems, and the ``vielbein'' of general relativity \cite{Padmanabhan10} for relating a coordinate basis to a non-coordinate unit basis.
While elasticity describes deformation of the same coordinate basis rather than a change of basis in the same configuration, the idea of treating one basis as an alternate non-coordinate non-unit basis seems an interesting possibility, although this would require introducing the concept of Cartan 
 connection to construct covariant derivatives.
   The use of anholonomic components of tensors was also discussed in the early mechanics literature \cite{EricksenTruesdell57}.
   I will also ignore the third, simple, Euclidean metric of the embedding space that induces the metrics of the reference and present configurations and is implicitly present when taking dot products. 
   More interesting background spaces are relevant to neutron stars \cite{CarterQuintana72}, and even to a more sophisticated treatment of incompatible elasticity \cite{OzakinYavari10}.
I note that in some solid mechanics literature, the present configuration is often loosely identified with the embedding space, and
working with present objects is sometimes unfortunately referred to as the ``Eulerian'' picture, despite the common use of ``Lagrangian'' convected coordinates and coordinate bases in the sense familiar to a fluid mechanician.

Another notable feature of some physics literature on elasticity is its inheritance of the use of geometric (per volume or per area) rather than material (per reference volume, akin to per mass) energies.
These geometric actions resemble those in high-energy and gravitational physics such as the Nambu-Goto action consisting of the area of the world-sheet of a ``string'', and the Einstein-Hilbert action consisting of the Ricci scalar integrated over the volume of space-time.  Despite names like ``string'' and ``brane'', such objects are geometric rather than elastic bodies, and the concept of material coordinates does not apply.
In the case of Einstein-Hilbert, there is no meaning to the position (embedding) vector either, and the variation of the volume form is performed with respect to the metric components, producing the Ricci scalar term in the associated field equations.
Similarly, the variation of a volume or area form with respect to position in a geometric energy will contribute a term proportional to the Lagrangian density.
There are physical situations where this is appropriate, but solid elasticity is not one of them, as it is described by a material energy defined per unit mass \cite{Peterson09}, an integral over a fixed set of material points.  
One expects a geometric energy for something like a soap film, where energy is proportional to surface area, and the film is able to pull material from a reservoir either at its boundaries, or within its own finite thickness.   
 In some situations, one might seek a shape that extremizes a functional without requiring it to describe a fixed amount of material. 
  This is quite distinct from the concept of virtual work in continuum mechanics, which relies on mass-conserving deformations \cite{GreenZerna92, Eringen67}.
How these issues might arise in attempts to construct variational principles in the context of growth \cite{BenAmarGoriely05, Yavari10, Nardinocchi13} is far beyond the scope of the current paper.

An energy commonly used to model lipid membranes as two-dimensional fluids with bending elasticity is the per area Helfrich energy \cite{Helfrich73, Evans74, Jenkins77jmb}. 
As this elasticity arises from different origins than does the bending elasticity of solid sheets, 
this paper does not 
cast any light on the appropriateness of such a choice. 
 The use of such an energy leads to geometrically elegant results \cite{Seifert97, CapovillaGuven02, Capovilla02, Guven04, Powers10, Deserno15}, and the advantage that terms arising from Gaussian curvature can be pushed out to the boundary conditions, or ignored in the case of closed membranes. 
Additionally, if the surface metric is constrained, differences in geometric and material energies can be absorbed into the corresponding multiplier, as shown in Section \ref{notgeometry}.
Otherwise, such differences might be justifiable as a small-strain approximation \cite{Leonard61, Koiter66, Dias11}. 
Steigmann \cite{AgrawalSteigmann09, Steigmann13ijnlm, Steigmann13ijnlmcorr} distinguishes between per area and per reference area energies, discusses how material coordinates are still relevant for fluid membranes, and comments on how physicists' conception of ``reparameterization'' may obscure the fact that variation of a material position vector means that actual material elements, not coordinates, are being moved around.

The remainder of the paper is broadly divided into the topics of bulk elasticity (Section \ref{elasticity}), reduction and related issues pertaining to bending energies (Section \ref{reduction}), and plate elasticity (Section \ref{plates}).  A few additional points and suggestions of further avenues for exploration are discussed in Section \ref{discussion}.
Notations, important concepts, and the definitions of fundamental geometric objects are provided in Section \ref{elasticitynotation}.
A few additional tools are introduced in later sections as needed to perform operations. 

The paper covers several major and minor topics.
The consequences of the ``choice of metric'' on the forms of both the bulk quadratic energy, in terms of invariants of Green or Almansi strain, and of its resulting equilibrium equations and boundary conditions, are presented across  
Sections \ref{elasticityreference}, \ref{elasticitypresent}, and \ref{elasticityhybrid}.  Some aspects of these derivations are interpreted in terms of standard continuum mechanical objects in  Appendix \ref{continuum}.
Emphasis is laid on the divergence form of the equations, both in these sections and in a later discussion of plate equations (Section \ref{platecombining}) that combine stretching or an in-plane metric constraint (Section \ref{platestretching}) with quadratic mean (Section \ref{platemeancurvature}) and Gaussian (Section \ref{plategaussiancurvature}) curvature elasticity terms arising from an Almansi form of the bulk energy.  Variation of these geometric quantities is presented in a compact vector form with intermediate steps made explicit.
There are significant differences in definitions of bending elasticity arising from the choice of these strain measures, or other possible choices, and in Section \ref{biotstrains}
it is suggested that the Biot strain may be a more appropriate choice for thin structures, an idea already present in prior literature but not in wide circulation
\cite{IwakumaKuranishi84, Chaisomphob86, Magnusson01, IrschikGerstmayr09, OshriDiamant16, OshriDiamant17}. 
These points are illustrated in the simple context of extensible \emph{elastica} in Section \ref{extensibleelastica}.
Section \ref{reductionplate} contains a brief sketch of reduction employing the Kirchhoff-Love assumptions to obtain a plate energy, and a discussion of some ambiguities in how terms are retained (Section \ref{reductionextraterms}).
The difference between elastic and geometric energies arises in Section \ref{notgeometry} with a discussion of a soap film energy,  
and the manner in which these differences need to be absorbed into multipliers in the inextensible limit. 
 The distinction is reiterated in the contexts of extensible 
 \emph{elastica} and plate bending energies in Section \ref{furthernotgeometry}.
Its most glaring consequences are the non-Helfrich form and additional tangential component of the contribution to the 
equations from the
 elastic squared mean curvature energy
(Section \ref{platemeancurvature}), 
and the fact that the variation of elastic Gaussian curvature energy is not a pure divergence and thus contributes to the Euler-Lagrange equations 
(Section \ref{plategaussiancurvature}).
A discussion of small-strain approximations in the combined plate equations and a brief comparison with inextensible \emph{elastica} can be found in Section \ref{platecombining}.

\section{elasticity}\label{elasticity}

In this section, after an extensive discussion of notation and fundamental objects for bulk and thin-body elasticity, I will compare several theories for an elastic energy ``quadratic'' in strain components, and discuss differences between elastic and geometric energies.

\subsection{Notation, coordinates, derivatives, bases, metrics, strains, invariants, and some geometric preliminaries}\label{elasticitynotation}

Let us take as fundamental objects the position vectors of present and reference configurations, their derivatives, and the inverses of these derivatives.  These will be used to construct tensorial or other quantities used in calculations.
A body is described using a position vector $\bR\left(\{x^i\}\right)$, $i \in \{1, 2, 3\}$, in  
$\mathbb{E}^3$, where $\{x^i\}$ are convected material coordinates.  
Derivatives will be denoted with subscripts, $\partial_i \equiv \tfrac{\partial}{\partial x^i}\,$.  
A variation $\delta$ will always mean a first order
 change resulting from the operation $\bR \to \bR + \delta\bR$ or, in later sections, its lower-dimensional analogue; no confusion should arise with the Kronecker delta $\delta^i_j$, which will always be written with indices.
The present configuration $\bR$ has the metric $g_{ij} = \partial_i\bR\cdot\partial_j\bR$ induced by the embedding space $\mathbb{E}^3$; the dot product will always mean the usual Euclidean operation.  We will also employ a reference metric $\bar{g}_{ij}$, which for the purposes of this paper can be imagined as derived from a reference configuration $\bar{\bR}$, $\bar{g}_{ij} = \partial_i\bar{\bR}\cdot\partial_j\bar{\bR}$.  
However, such an object is often used for systems without a reference configuration in Euclidean space, as might be appropriate to describe thermal expansion or chemical swelling in frustrated incompatible-elastic bodies that cannot ``rest'' peacefully  \cite{Efrati09jmps}.  The present and reference metrics have inverses, such that $g^{ij}g_{jk} = \delta^i_k$ and $\left(\bar{g}^{-1}\right)^{ij}\bar{g}_{jk} = \delta^i_k$.  Instead of the latter expression, we will often write the same functions as $\bar{g}^{IJ}\bar{g}_{jk}=\delta^I_k$, with summation holding independent of capitalization, using capital raised indices to indicate raising with the inverse metric rather than the present metric.  
Thus the functions $\left(\bar{g}^{-1}\right)^{ij} = \bar{g}^{IJ} = \bar{g}^{IL}\bar{g}^{JK}\bar{g}_{LK}$ are distinct from the functions $\bar{g}^{ij} = g^{il}g^{jk}\bar{g}_{jk}$.
The present and reference configurations have as natural coordinate basis vectors their tangent vectors $\partial_i\bR$ and $\partial_i\bar{\bR}$, respectively.  The corresponding reciprocal basis vectors are $\partial^i\bR = g^{ij}\partial_j\bR$ and $\left(\partial\bar{\bR}^{-1}\right)^i = \left(\bar{g}^{-1}\right)^{ij}\partial_j\bar{\bR} = \partial^I\bar{\bR} = \bar{g}^{IJ}\partial_j\bar{\bR}$.  Note that $g^{ij} = \partial^i\bR\cdot\partial^j\bR$ and  $\bar{g}^{IJ} = \partial^I\bar{\bR}\cdot\partial^J\bar{\bR}$.  The deformation gradient can be written as $\partial_i\bR\partial^I\bar{\bR}$.

A great potential source of confusion in elasticity is the fact that these functions, and others derived from them, can serve as covariant or contravariant components of \emph{different tensors} by pairing them with either the present or reference bases.\footnote{If one works only with coordinate bases, this is equivalent to considering them as objects in the present or reference configurations.} 
 This is despite the fact that the indices always correspond to the same material coordinates--- the reference basis is convected into the present basis by the deformation.  It may be confusing if one is used to thinking of metrics in terms of line elements, as the expressions $dl^2 = g_{ij}dx^idx^j$ and $d\bar{l}^2 = \bar{g}_{ij}dx^idx^j$ involve the same material coordinate segments $dx^i$ that are unchanged by deformations of the body \cite{Peterson09}.
But the change of basis matters.
Unlike many other situations in physics, in elasticity one cannot unambiguously refer to a tensor by referring to its components. 

The two notations I adopt here avoid potential ambiguities associated with the otherwise very compact and elegant formalism of Green and Zerna \cite{GreenZerna92} that has been adopted with modifications by Steigmann \cite{Steigmann90} and others.
They also obviate any need to remember the respective spaces in which different objects live.
  The clumsy inverse notation is necessary when one writes out specific components, such as $\left(g^{-1}\right)^{12}$, and will also be of conceptual help in Section \ref{elasticityhybrid} when both bases are to be used simultaneously.
 Interesting alternate approaches to notation may be found in \cite{NguyenSelinger17, vanRees17}. 

Paired with the (reciprocal) reference basis, the $g_{ij} = g_{IJ}$ are covariant components of the right Cauchy-Green deformation tensor $\bC = g_{IJ}\partial^I\bar{\bR}\partial^J\bar{\bR}$, while paired with the (reciprocal) present basis they are covariant components of the identity $\bI = g_{ij}\partial^i\bR\partial^j\bR$ and colloquially said to be ``the metric'', as it is implied that one is in the present configuration.
Similarly, the $\bar{g}_{ij}=\bar{g}_{IJ}$ are either covariant components of the identity $\bI = \bar{g}_{IJ}\partial^I\bar{\bR}\partial^J\bar{\bR}$, in which case they are ``the metric'' in the reference configuration,\footnote{Both the present and reference configurations of a three-dimensional body inherit the metric of $\mathbb{E}^3$, which is simply the identity.}
 or they are covariant components of the inverse left Cauchy-Green tensor $\bB^{-1} = \bar{g}_{ij}\partial^i\bR\partial^j\bR$.
For completeness and later use, let us also write out the corresponding inverse right Cauchy-Green tensor $\bC^{-1} = g^{ij}\partial_i\bar{\bR}\partial_j\bar{\bR}$ and the left Cauchy-Green tensor $\bB = \bar{g}^{IJ}\partial_I\bR\partial_J\bR$.  Just as with the metric functions, capitalization only matters on the upper indices, so $\partial_i\bR = \partial_I\bR$ and $\partial_i\bar{\bR}=\partial_I\bar{\bR}$, and the important distinguishers are the bars or lack thereof on the $\bR$s.  This is simply a consequence of the fact that $\partial_i$ and $\partial_I$ are the same partial derivative, but more care will be needed when we employ covariant derivatives.
 
An important derived quantity is the difference in two metrics
\begin{align}
 	\epsilon_{ij} = \epsilon_{IJ} = \tfrac{1}{2}\left( g_{ij} - \bar{g}_{ij} \right) \, , \label{epsilon}
\end{align}
 which is a commonly used measure of strain.  These functions are, again, components of two possible tensors: 
the Green(-Lagrange-St.Venant) strain tensor $\bm{\epsilon}_{\text{\tiny{G}}} = \epsilon_{IJ}\partial^I\bar{\bR}\partial^J\bar{\bR} = \tfrac{1}{2}\left(\bC - \bI \right)$ or the \mbox{(Euler-)Almansi} strain tensor $\bm{\epsilon}_{\text{\tiny{A}}} = \epsilon_{ij}\partial^i\bR\partial^j\bR = \tfrac{1}{2}\left(\bI - \bB^{-1}\right)$.
With the bases in mind, it is clear how to form invariants of such tensors from their identical covariant components.  For example, the traces of the Green and Almansi strains have the unambiguous definitions $\mathrm{Tr}( \bm{\epsilon}_{\text{\tiny{G}}} ) = \bI : \bm{\epsilon}_{\text{\tiny{G}}} = \bar{g}^{IJ}\epsilon_{IJ} = \left(\bar{g}^{-1}\right)^{ij}\epsilon_{ij}$ and $\mathrm{Tr}( \bm{\epsilon}_{\text{\tiny{A}}} ) = \bI : \bm{\epsilon}_{\text{\tiny{A}}} = g^{ij}\epsilon_{ij}$, respectively.  These are sometimes informally expressed as $\mathrm{Tr}_{\bar{g}}(\epsilon)$ and $\mathrm{Tr}_{g}(\epsilon)$, or by saying that one takes the trace of the strain by raising with one (inverse) metric or the other, but really this doesn't have a meaning.  In elasticity, the component functions do not unambiguously define a tensor, so the same functions can have different interpretations.
Thus the question of which metric is used to define invariants is really a question of which energy or constitutive relation is being chosen, and this does not have a ``right'' answer except as it connects to experiments on particular materials.

Each metric has its own compatible covariant derivative identifiable with its Levi-Civita connection, such that $\bar{\nabla}_K\bar{g}_{IJ} = 0$ and $\nabla_k g_{ij} = 0$, with corresponding Christoffel symbols formed with the respective metrics and inverse metrics.
Likewise the metric determinants pass through the corresponding derivatives, $\bar{\nabla}_K\sqrt{\bar{g}}=0$ and $\nabla_k\sqrt{g}=0$.  
A single partial derivative of a vector such as $\bR$, with no indices, may be replaced by a material covariant derivative, $\partial_i\bR = \nabla_i\bR = \bar{\nabla}_I\bR$, an object with one covariant material index which should subsequently be interpreted carefully depending on context.
Another property of $\mathbb{E}^3$ is that $\bar{\nabla}_I \bar{\nabla}_J \bar{\bR} = {\bf{0}}$ and $\nabla_i\nabla_j\bR = {\bf{0}}$, so we could define displacements $\bu = \bR - \bar{\bR}$ and write $\bar{\nabla}_I\bar{\nabla}_J \bu = \bar{\nabla}_I\bar{\nabla}_J \bR$ and $\nabla_i\nabla_j \bu = -\nabla_i\nabla_j\bar{\bR}$.
Note that
\begin{align}
	\partial_i\partial_j\bR &= \partial_i\nabla_j\bR = \partial_i\partial_j\bR\cdot\nabla^k\bR\nabla_k\bR = \Gamma^k_{ji}\partial_k\bR \, , \\
	\mathrm{so}\quad \nabla_i\nabla_j\bR &= \partial_i\nabla_j\bR - \Gamma^k_{ij}\nabla_k\bR = {\bf{0}} \, , 
\end{align}
using the fact that the Levi-Civita connection is torsion-free, so the Christoffel symbols $\Gamma^k_{ij}$ are symmetric in their lower indices. 
However, when we consider plates, we will use a two-dimensional description of the mid-surface of the body, with only the surface indices as material indices. 
Let $\bX\left(\{x^\alpha\}\right)$, $\alpha \in \{1, 2\}$, be a two-dimensional body, a surface embedded in $\mathbb{E}^3$, with unit normal $\uvc{N}$, metric $a_{\alpha\beta} = \partial_\alpha\bX\cdot\partial_\beta\bX$, and symmetric curvature tensor components  $b_{\alpha\beta} = \partial_\beta\partial_\alpha\bX\cdot\uvc{N} = {-\partial_\alpha\bX\cdot\partial_\beta\uvc{N}}$.  It is also convenient to define the components of the third fundamental form ${\partial_\alpha\uvc{N}\cdot\partial_\beta\uvc{N}} = b^\gamma_\alpha b_{\beta\gamma} = c_{\alpha\beta}$.
Now
 \begin{align}
	\partial_\alpha\partial_\beta\bX &= \partial_\alpha\nabla_\beta\bX = \partial_\alpha\partial_\beta\bX\cdot\left( \nabla^\gamma\bX\nabla_\gamma\bX + \uvc{N}\uvc{N} \right)	 = \Gamma^\gamma_{\beta\alpha}\partial_\gamma\bX + b_{\beta\alpha}\uvc{N} \, , 
\end{align}
and as $\nabla_\alpha\nabla_\beta\bX = \partial_\alpha\nabla_\beta\bX - \Gamma^\gamma_{\alpha\beta}\nabla_\gamma\bX$, we have obtained the first part of the Gauss-Weingarten system
\begin{align}
	\nabla_\beta\nabla_\alpha\bX &= b_{\alpha\beta}\uvc{N} \, , \label{GW1} \\
	\nabla_\alpha\uvc{N} &= -b^\beta_\alpha\nabla_\beta\bX \, . \label{GW2}
\end{align}
Applied to the position vector of the surface, two covariant derivatives give a term, symmetric in its two surface indices, that does not vanish but points entirely normal to, or rather out of, this lower-dimensional curved material space.  
 Note that this is distinct from pointing normal to the boundary of the body--- there is no corresponding concept for a space-filling body. 
More broadly, while covariant derivatives do not in general commute in a curved space, they will do so when acting on a scalar or something like a spatial vector $\bX$ that carries no material indices.

The mean and Gaussian curvatures of $\bX$ are invariants of the curvature tensor $\bb = b_{\alpha\beta}\partial^\alpha\bX\partial^\beta\bX$,
\begin{align}
	H &= \tfrac{1}{2}b^\alpha_\alpha = \tfrac{1}{2}\, \mathrm{Tr}\left(\bb\right) \, , \\
	K &= \tfrac{1}{2}\left( b^\alpha_\alpha b^\beta_\beta - b^\alpha_\beta b^\beta_\alpha \right) = \tfrac{1}{2}\,\left( \left(\mathrm{Tr}\left(\bb\right)\right)^2 -\mathrm{Tr}\left(\bb^2\right) \right) =\mathrm{Det}\left(\bb\right)\, ,
\end{align}
where clearly 
the inverse metric of the surface $a^{\alpha\beta}$ is used to form invariants from derivatives of position.  
Notation such as that in \cite{Armon10} and \cite{Pezzulla17} is vague, as it suggests that a curvature tensor is being used when it is not.
The quadratic invariants relevant to bending energies can also be written directly in terms of the derivatives,
\begin{align}
	H^2 &= \tfrac{1}{4}\nabla^2\bX\cdot\nabla^2\bX \, , \label{meanX} \\
	K &= \tfrac{1}{2}\nabla^2\bX\cdot\nabla^2\bX - \tfrac{1}{2}\nabla^\alpha\nabla_\beta\bX\cdot\nabla_\alpha\nabla^\beta\bX \, , \label{GaussX}
\end{align}
where $\nabla^2$ stands for the covariant Laplacian or Laplace-Beltrami operator on the surface $\nabla^\alpha\nabla_\alpha$.

Finally, the compatibility relationships between metric and curvature for two-dimensional surfaces are contained in the Codazzi equations
\begin{align}
	\nabla_\alpha b_{\beta\gamma} = \nabla_\beta b_{\alpha\gamma} \, , \label{Codazzi}
\end{align}
and the Gauss equation
\begin{align}
	b^\alpha_\alpha b^\beta_\beta - b^\alpha_\beta b^\beta_\alpha = 2K  &= R\indices{^\alpha^\beta_\alpha_\beta} = R\indices{^\alpha_\alpha} = R \, , \label{Gauss} \\
	\mathrm{or\,\, equivalently\quad} K\delta^\alpha_\beta  &= R\indices{^\gamma^\alpha_\gamma_\beta} = R\indices{^\alpha_\beta} \, , \label{KR}
\end{align}
where $R_{\alpha\beta\gamma\zeta}$ is the Riemann tensor, definable in terms of the metric \emph{via} the usual mess of Christoffel symbols, $R_{\alpha\beta}$ is the Ricci tensor, and $R$ is the Ricci scalar. 

\subsection{Reference metric as \emph{the} metric: Invariants of Green strain in the reference configuration}\label{elasticityreference}

This approach, in which the reference metric plays every metric role, is closest to the influential work of Efrati and co-workers \cite{Efrati09jmps} 
which has been adopted by several other groups. 
While not particularly traditional, there is precedent for this approach in the shell literature \cite{Budiansky68}.
In Section \ref{extensibleelastica}, we will demonstrate that the bending energy for thin bodies derived from this bulk energy has undesirable features.
I tentatively interpret what follows as performing all operations in the reference configuration.
In Section \ref{elasticityhybrid}, we will revisit the same energy in the present configuration.  Appendix \ref{continuum} will provide further clarification of these two treatments.

We write the energy as 
\begin{align}
	\bar{E} = \int d\bar{V} \bar{\mathcal{E}} \, , \label{energyref}
\end{align}
where $d\bar{V} = \sqrt{\bar{g}} \prod\limits_{i} dx^i$. 
The energy density $\bar{\mathcal{E}}$ can be thought of as per unit unstrained (reference) volume, or equivalently as per unit mass if we assume a uniform mass density of unity in the reference configuration.  
For simplicity, the distinction will be ignored in this paper, and the phrase ``per unit mass'' used loosely.  However, it is important in more general problems involving a combination of swelling and subsequent elastic deformation in which multiple reference states exist \cite{NguyenSelinger17}.

The strain components in \eqref{epsilon} are taken as the relevant measure of strain.  This is by no means the only or best choice, as will be discussed in Section \ref{biotstrains}.  However, given this choice, the appropriate energy density for small strains is the quadratic expression
\begin{align}
	\bar{\mathcal{E}} = \tfrac{1}{2}\bar{A}^{IJKL}\epsilon_{IJ}\epsilon_{KL} \, ,\label{energydensityref}
\end{align}
where the isotropic elasticity tensor has contravariant components\\ $\bar{A}^{IJKL} = \lambda \bar{g}^{IJ}\bar{g}^{KL} + \mu\left(\bar{g}^{IK}\bar{g}^{JL} + \bar{g}^{IL}\bar{g}^{JK}\right)$, $\lambda$ and $\mu$ being constant moduli with units of energy per unit mass.  
  Thus $2\bar{\mathcal{E}} = \lambda\epsilon_I^I\epsilon_J^J + 2\mu\epsilon_J^I\epsilon_I^J = \lambda\left(\mathrm{Tr}\left(\bm{\epsilon}_{\text{\tiny{G}}}\right)\right)^2 + 2\mu\mathrm{Tr}\left(\bm{\epsilon}_{\text{\tiny{G}}}^2\right)$.

The variation $\bR \to \bR + \delta\bR$ passes through the partial derivative, and transforms the present metric as $\delta g_{ij} = \partial_i\delta\bR\cdot\partial_j\bR + \partial_i\bR\cdot\partial_j\delta\bR$.  It leaves $\bar{g}_{ij}$, $\bar{g}^I_j = \delta^I_j$, and $\bar{g}^{IJ}$ unchanged, but transforms $g^I_j = \bar{g}^{IK}g_{kj}$.
To perform the variation, we use the equivalence of $\partial_i\bR = \bar{\nabla}_i\bR$, the $(IJ,KL)$ symmetry of $A^{IJKL}$, 
and the following fact. Let there be a symmetric tensor with components $S_{ij} = Q_{ij} + Q_{ji}$.  Then $S^i_iS^j_j = 4Q^i_iQ^j_j$ and $S^i_jS^j_i = 4Q^i_jQ^j_i$.  Thus,
\begin{align}
	\delta\bar{\mathcal{E}} &= \delta \int d\bar{V} \tfrac{1}{8}\bar{A}^{IJKL}\left(g_{IJ}-\bar{g}_{IJ}\right)\left(g_{KL}-\bar{g}_{KL}\right) \, , \nonumber \\
	&= \int d\bar{V} \tfrac{1}{2}\bar{A}^{IJKL}\left(g_{IJ}-\bar{g}_{IJ}\right) \bar{\nabla}_L\bR\cdot\bar{\nabla}_K\delta\bR  \, , \nonumber \\
	&= -\int d\bar{V}\, \bar{\nabla}_K\left( \bar{A}^{IJKL}\epsilon_{IJ} \bar{\nabla}_L\bR \right) \cdot \delta\bR \, \nonumber \\
	&\quad\,\, +	\oint d\bar{S}\, \bar{n}_K \bar{A}^{IJKL}\epsilon_{IJ} \bar{\nabla}_L\bR \cdot \delta\bR \, ,
\end{align}
assuming a smooth boundary $\partial \bar{V}$.   Here $d\bar{S}$ is a surface element constructed with the reference metric and $\bar{n}_K$ is the covariant component, in the reference basis, of the unit normal to the reference configuration, $\bm{\mathrm{\hat{\bar{n}}}} = \bar{n}^I\partial_I\bar{\bR}$ and $\bm{\mathrm{\hat{\bar{n}}}}\cdot\partial_J\bar{\bR} = \bar{n}_J$.

Thus we are led to define stress components
\begin{align}
	\bar{\Sigma}^{KL} = \bar{A}^{IJKL}\epsilon_{IJ} \, , \label{stressref}
\end{align}
such that the equations of equilibrium are
\begin{align}
	-\bar{\nabla}_K\left(\bar{\Sigma}^{KL} \bar{\nabla}_L\bR \right) = {\bf{0}} \, , \label{bulkref}
\end{align}
and free boundary conditions on $\partial \bar{V}$ are
\begin{align}
	\bar{n}_K \bar{\Sigma}^{KL} \bar{\nabla}_L\bR  = {\bf{0}} \, . \label{bcref}
\end{align}
These results are equivalent to those in \cite{GreenZerna92, Budiansky68,
Steigmann90}, and differ from those of \cite{Efrati09jmps} only in that their expressions use the present normal vector.

It seems strange that the basis vectors of the present configuration $\bar{\nabla}_L\bR = \partial_l\bR$ are here playing the role of covariant components of objects in the reference configuration.  I will revisit this thought in Section \ref{elasticityhybrid}.

Throughout the paper, I will informally interpret \eqref{bulkref} and other expressions like it as being vectors in the embedding space and scalars in the material space of the body. 
  It is apparent that \eqref{bulkref} is a divergence, as it should be.
However, $\bar{\nabla}_K\bar{\nabla}_L \bR \ne 0$ because the reference metric and connection are not constructed from the present configuration $\bR$.  For this reason, the projection (by Euclidean dot product) of the divergence \eqref{bulkref} onto the tangent vectors does \emph{not} result in a divergence form of the resulting equations for material 
components.
Indeed the projection onto the inverse deformed tangent vector $\bar{\nabla}^J\bR$ of the bulk equation \eqref{bulkref} is cumbersome,
\begin{align}
	-g_L^J \bar{\nabla}_K\bar{\Sigma}^{KL} -\bar{\Sigma}^{KL}\bar{\nabla}_K\bar{\nabla}_L\bR\cdot\bar{\nabla}^J\bR    = 0 \, . \label{bulkrefproj}
\end{align}
This is accompanied by the projection of the boundary conditions \eqref{bcref}
\begin{align}
	g_L^J \bar{n}_K\bar{\Sigma}^{KL}   = 0 \, . \label{bcrefproj}
\end{align}
We may abusively rewrite the second derivatives as
\begin{align}
	\bar{\nabla}_K\bar{\nabla}_L\bR &= \nabla_K\nabla_L\bR + \left(\Gamma^m_{KL} - \bar{\Gamma}^M_{KL}\right)\partial_M\bR \, , 
\end{align}
where Christoffel symbols are defined using the corresponding unbarred or barred metrics, and upper indices raised following our capitalization convention.  
We may also eliminate stray mixed components of present metrics by returning to  \eqref{bulkref} and \eqref{bcref} and projecting them onto 
 $\bar{\nabla}^J\bR$,
 to obtain the projection of the bulk equation
\begin{align}
		-\bar{\nabla}_K\bar{\Sigma}^{KJ} - \bar{\Sigma}^{KL}\left(\Gamma^j_{KL} - \bar{\Gamma}^J_{KL}\right) = 0 \, , \label{bulkrefproj2}
\end{align}
with boundary conditions
\begin{align}
	\bar{n}_K\bar{\Sigma}^{KJ}   = 0 \, . \label{bcrefproj2}
\end{align}
Note that \eqref{bulkrefproj2} is typically written with some type of explicit inverse notation, akin to what was introduced in Section \ref{elasticitynotation} but with metric roles reversed.
I think we need to interpret the free index as corresponding to the same configuration for each term; this is likely the reference configuration, despite the fact that in one term the index was raised with the present metric.  
The representations \eqref{bulkref} or \eqref{bulkrefproj} for the bulk equation are less ambiguous.
Equations (\ref{bulkrefproj2}-\ref{bcrefproj2}) can be found in \cite{Efrati09jmps}, excepting notational differences and the disagreement of the choice of normal.  
Similar expressions appear in Sanders \cite{Sanders63}.
The difference in Christoffel symbols is a tensor that also appears in other works \cite{John65, Steigmann12, KarloviniSamuelsson03}.  
Steigmann \cite{Steigmann12} refers to it as a ``strain measure'' for shells, along with more typical strain and curvature differences.  

Because $\bar{\nabla}_K\bar{\nabla}_L \bR$ is $O(\epsilon, \partial_i\epsilon)$ and $g^J_L = \delta^J_L + O(\epsilon)$, it may be quite sensible to write approximate projected bulk equations
\begin{align}
	-\bar{\nabla}_K\bar{\Sigma}^{KJ} + O(\epsilon^2, \epsilon\partial_i\epsilon) = 0 \, .
\end{align}
There seems little reason to include quadratic terms in the Euler-Lagrange equations while neglecting others that would have been generated by an energy expanded out to cubic order.
However, in Section \ref{elasticityhybrid} I will present a better way to clean up these expressions that does not rely on such an argument, and naturally provides a divergence form for both the equations and their projections.

\subsection{Present metric as \emph{the} metric: Invariants of Almansi strain in the present configuration}\label{elasticitypresent}

This approach, in which the present metric plays every metric role, is a somewhat unconventional constitutive choice in continuum mechanics. 
However, this is essentially the default 
among physicists studying relativistic elasticity or the bending elasticity of fluid membranes (whether biological or cosmic), due in part to its natural production of geometric quantities in the equations.
Such an approach is often applied in a manner inappropriate for describing a fixed quantity of elastic material; the distinction between geometric and material energies will be discussed in Section \ref{notgeometry}.
The particular presentation here, as well as its further application to plates in Section \ref{plates} is, I believe, not to be found elsewhere.
In Section \ref{extensibleelastica}, we will see that the qualitative properties of the bending energy derived from this approach are in direct opposition to those derived from the reference metric approach, and that neither correspond to the simplest director theories of rods.
All operations are performed in the present configuration.

We write the energy as
\begin{align}
	E = \int dV \mathcal{E} \label{energypresent} \, ,
\end{align}
where $dV = \sqrt{g} \prod\limits_{i} dx^i = \sqrt{g/\bar{g}}\,d\bar{V}$. 
This description now requires the density $\mathcal{E}$ to be an energy per unit strained (present) volume.
Yet we still expect the moduli to be constants per unit mass, not per unit volume.  Thus a sensible energy density is 
\begin{align}
	\mathcal{E} = \tfrac{1}{2}\sqrt{\bar{g}/g}\,A^{ijkl}\epsilon_{ij}\epsilon_{kl} \, ,\label{energydensitypresent}
\end{align}
where the isotropic elasticity tensor has contravariant components $A^{ijkl} = \lambda g^{ij}g^{kl} + \mu\left(g^{ik}g^{jl} + g^{il}g^{jk}\right)$.
Thus $2\mathcal{E} = \sqrt{\bar{g}/g}\left(\lambda\epsilon_i^i\epsilon_j^j + 2\mu\epsilon_j^i\epsilon_i^j \right) = \sqrt{\bar{g}/g}\left(\lambda\left(\mathrm{Tr}\left(\bm{\epsilon}_{\text{\tiny{A}}}\right)\right)^2 + 2\mu\mathrm{Tr}\left(\bm{\epsilon}_{\text{\tiny{A}}}^2\right)\right)$.
Note that while $\sqrt{\bar{g}}$ and $\sqrt{g}$ are scalar densities, a quantity like the scale factor $\sqrt{\bar{g}/g}$ is a scalar because the two metric determinants transform equally under changes of the same material coordinates.  The quantities $\sqrt{g/\bar{g}}$ and $\sqrt{\bar{g}/g}$ usually appear as $J$ and $J^{-1}$ in the continuum mechanics literature.
For the purposes of the variation, the $\sqrt{g}$ in the volume form $dV$ and the $1/\sqrt{g}$ multiplying the moduli cancel each other out.  Physically this just means that the energy for a fixed amount of material does not increase simply because the volume increases.  This is a material energy, not a geometric energy like that of a soap film in contact with a reservoir (Section \ref{notgeometry}).

Although they look similar, $\mathcal{E} \ne \bar{\mathcal{E}}$ and $E \ne \bar{E}$, and these approaches represent two different theories of elasticity arising from two different constitutive stored energy functions.  Either could be said to be ``quadratic'' in the strain components, but that is not enough to uniquely specify the theory, as the components can belong to either the Green or Almansi strain tensors.
The distinction is of higher order than quadratic in strain, that is, 
\begin{align}
	\mathcal{E} &= \bar{\mathcal{E}} + O(\epsilon^3) \, , \label{compare}
\end{align}
and indeed this is also the argument for the rather unphysical neglect of $\sqrt{\bar{g}/g}$ in \cite{Dias11}.
However, as we will see in 
Sections \ref{reductionextraterms} and \ref{plateextraterms}, such differences in energy actually have important lower-order consequences at the level of the equations of equilibrium for thin bodies, which might be an artifact of these choices of strain measure.
To see that \eqref{compare} is true, note that the elasticity tensor used in the reference energy in Section \ref{elasticityreference} would be written in the present configuration using components of the inverse reference metric $\left(\bar{g}^{-1}\right)^{ij} = \bar{g}^{IJ}$.  We can explicitly relate the mixed components of the inverse reference metric and those of the present metric $g^i_k = \delta^i_k$:
\begin{align}
	2\left(\bar{g}^{-1}\right)^i_j\epsilon^j_k &= \left(\bar{g}^{-1}\right)^i_j\left(\delta^i_k - \bar{g}^j_k\right) \, , \nonumber \\
	&= \left(\bar{g}^{-1}\right)^i_k - \delta^i_k \, .
\end{align}
Additional examples of such approximations appear in the early shell literature. Koiter \cite{Koiter66} invokes small strain arguments to replace the deformed surface metric tensor with the undeformed metric tensor in the compatibility equations for strains, thus only approximately satisfying Gauss and Codazzi.  He also approximates the relevant scale factor as unity in writing a quadratic energy.  Leonard \cite{Leonard61} also discusses similar substitutions of metrics for small strain theories.  Relationships between the quadratic and cubic invariants of the relevant tensors have been known since the early days of nonlinear elasticity \cite{DoyleEricksen56, GreenZerna92}.

To perform the variation $\bR \to \bR + \delta\bR$, we will need 
\begin{align}
	\delta g^{ij} &= -a^{ik}a^{jl}\delta g_{kl} \, , \nonumber \\
	&= -\nabla^i\bR\cdot\nabla^j\delta\bR - \nabla^j\bR\cdot\nabla^i\delta\bR \, , \label{varinversemetric3D}
\end{align}
where we have used $\delta\left(g^{jl}g_{lk}\right) = 0$, the symmetry of $g_{ij}$, and the equivalence $\partial_i\bR = \nabla_i\bR$.  
The variation still leaves the reference metric components $\bar{g}_{ij}$ and the delta functions $g^i_j = \delta^i_j$ unchanged, but $\delta\bar{g}^i_j = \delta g^{ik} \bar{g}_{kj}$, and similarly $\delta\bar{g}^{ij} \ne 0$. 
As mentioned above, there is no need to consider a contribution to $\delta\mathcal{E}$ from the variation of the volume form $dV$, as this is compensated by the prefactor $\sqrt{\bar{g}/g}$ in the energy density: $\delta\left(dV\sqrt{\bar{g}/g}\right) = 0$.
Using the same tricks as before, the variation is most easily performed thus:
\begin{align}
	\delta E &= \delta \int dV \tfrac{1}{8}\sqrt{\bar{g}/g}\,A^{ijkl}\left(g_{ij}-\bar{g}_{ij}\right)\left(g_{kl}-\bar{g}_{kl}\right) \, , \nonumber \\
	&= \int dV \tfrac{1}{2}\sqrt{\bar{g}/g}\,A\indices{_i^j_k^l}\left(\delta^i_j-\bar{g}^i_j\right) \bar{g}_{ml}\nabla^m\bR\cdot\nabla^k\delta\bR \, , \nonumber \\
	&= -\int dV\, \nabla_k\left( \sqrt{\bar{g}/g}\,A^{ijkl}\epsilon_{ij} \bar{g}^m_l\nabla_m\bR \right) \cdot \delta\bR \, \nonumber \\
	&\quad\,\, +	\oint dS\, \sqrt{\bar{g}/g}\,n_kA^{ijkl}\epsilon_{ij} \bar{g}^m_l\nabla_m\bR \cdot \delta\bR \, ,
\end{align}
assuming a smooth boundary $\partial V$.  Here $dS$ is a surface element constructed with the present metric and $n_k = \uvc{n}\cdot\partial_k\bR$ is the covariant component, in the present basis, of the unit normal to the present configuration.
 
Thus we define stress components
\begin{align}
	\Sigma^{km} &= A^{ijkl}\epsilon_{ij} \bar{g}^m_l \, , \label{stresspresent} \\
	&= A^{ijkm}\epsilon_{ij} - 2A^{ijkl}\epsilon_{ij}\epsilon^m_l \, ,
\end{align}
such that the equations of equilibrium are
\begin{align}
	-\nabla_k\left( \sqrt{\bar{g}/g}\, \Sigma^{km} \nabla_m\bR \right) = {\bf{0}} \, , \label{bulk}
\end{align}
and free boundary conditions on $\partial V$ are
\begin{align}
 	\sqrt{\bar{g}/g}\, n_k \Sigma^{km} \nabla_m\bR  = {\bf{0}} \, . \label{bc}
\end{align}
A nice thing about the divergence \eqref{bulk} is that its projection (onto $\nabla^j\bR$) is also 
 a divergence,
\begin{align}
	-\nabla_k\left(\sqrt{\bar{g}/g}\, \Sigma^{kj}\right) = 0 \, , \label{bulkproj}
\end{align}
because $\nabla_k\nabla_m\bR = {\bf{0}}$. 
Similarly, we have the boundary conditions
\begin{align}
	\sqrt{\bar{g}/g}\, n_k \Sigma^{kj}  = 0 \, . \label{bcproj}
\end{align}
These results are equivalent to those in \cite{GreenZerna92}.

Again, as only quadratic strain terms were retained in the energy, it may be sensible to  ignore some of the quadratic strain terms in the definition of the stress components,
\begin{align}
	\Sigma^{kl} = A^{ijkl}\epsilon_{ij} + O(\epsilon^2) \, . \label{stresspresentapprox}
\end{align}
Had we followed the more common practice of constructing the invariants with the reference metric, or rather using the invariants of the Green strain,  
 the only difference in our result would have been a quadratic adjustment to the stress.
For now we retain, rather than approximate, the prefactor $\sqrt{\bar{g}/g}\,$ appearing in front of the stress, both to maintain some similarity with the continuum mechanics literature, and because it has a sensible physical meaning connected to the conservation of mass.
Also, while $\sqrt{g}$ passes through its connection $\nabla$, it is simplest to keep the scalar factor $\sqrt{\bar{g}/g}$ intact, because $\sqrt{\bar{g}}$ is a tensor density and does not behave as a scalar under covariant differentiation.  That is, 
\begin{align}
	\nabla_i\sqrt{\bar{g}} &= \sqrt{g}\,\nabla_i\sqrt{\bar{g}/g} \, , \nonumber \\
	&= \sqrt{g}\, \partial_i\sqrt{\bar{g}/g} \, , \nonumber \\
	&= \partial_i\sqrt{\bar{g}}\, - \sqrt{\bar{g}/g}\, \partial_i \sqrt{g} \, , \nonumber \\
	&= \partial_i\sqrt{\bar{g}}\, - \sqrt{\bar{g}}\,\Gamma^j_{ji} \, ,
\end{align}
and therefore even a simple Cartesian reference integration measure will not have a simple present covariant derivative.  It is easier to deal with $\nabla_i\sqrt{\bar{g}/g} = \partial_i\sqrt{\bar{g}/g}\,$.  Of course, it is even easier to approximate $\sqrt{\bar{g}/g}$ as unity, and throw away the terms of higher order in strain.

\subsection{Hybrid approach: Invariants of Green (or rather some other unnamed) strain in the present configuration }\label{elasticityhybrid}

This approach is somewhat more traditional than that of either of the two previous sections, but requires some care in interpretation.
Let us work with the present metric, but rewrite the reference energy \eqref{energyref} as an integral over the present configuration
\begin{align}
	\bar{E} = \int dV \sqrt{\bar{g}/g}\, \bar{\mathcal{E}} \, . \label{energyrefinpresent}
\end{align}
We define $\bar{\mathcal{E}} = \tfrac{1}{2}\bar{A}^{ijkl}\epsilon_{ij}\epsilon_{kl}$ using the same functions as in \eqref{energydensityref}, the difference being that we now consider the components of the elasticity tensor as referring to the present basis, even though they are computed using the inverse reference basis.  We write $\bar{A}^{ijkl} = \lambda \left(\bar{g}^{-1}\right)^{ij}\left(\bar{g}^{-1}\right)^{kl} + \mu\left(\left(\bar{g}^{-1}\right)^{ik}\left(\bar{g}^{-1}\right)^{jl} + \left(\bar{g}^{-1}\right)^{il}\left(\bar{g}^{-1}\right)^{jk}\right)$, and generate the same quantities as in Section \ref{elasticityreference}.
We are obliquely using the fact that we can view the same component functions in two ways, and obtain the same invariants from the two different tensors $\bar{g}^{IJ}\partial_I\bar{\bR}\partial_J\bar{\bR} \cdot 2\epsilon_{KL}\partial^K\bar{\bR}\partial^L\bar{\bR} = \bI \cdot \left(\bC - \bI\right) = \bC - \bI$ and 
$\left(\bar{g}^{-1}\right)^{ij}\partial_i\bR\partial_j\bR \cdot 2\epsilon_{kl}\partial^k\bR\partial^l\bR = \bB \cdot \left(\bI - \bB^{-1}\right) = \bB - \bI$.
While perhaps a bit confusing at first, this seems easier to interpret than the reference metric approach in the context of incompatible elasticity, when no reference basis exists.\footnote{It may seem hard to believe that there are a few animals in the continuum mechanical zoo without one or more standard names, but the tensor $\bB - \bI$ appears to be one of them.}
Proceeding with the variation, we have
\begin{align}
	\delta\bar{\mathcal{E}} &= \delta \int dV \tfrac{1}{8}\sqrt{\bar{g}/g}\,\bar{A}^{ijkl}\left(g_{ij}-\bar{g}_{ij}\right)\left(g_{kl}-\bar{g}_{kl}\right) \, , \nonumber	 \\
	&= \int dV \tfrac{1}{2}\sqrt{\bar{g}/g}\,\bar{A}^{ijkl}\left(g_{ij}-\bar{g}_{ij}\right) \nabla_l\bR\cdot\nabla_k\delta\bR  \, , \nonumber \\
	&= -\int dV\, \nabla_k\left( \sqrt{\bar{g}/g}\,\bar{A}^{ijkl}\epsilon_{ij} \nabla_l\bR \right) \cdot \delta\bR \, \nonumber \\
	&\quad\,\, +	\oint dS\, \sqrt{\bar{g}/g}\, n_k \bar{A}^{ijkl}\epsilon_{ij} \nabla_l\bR \cdot \delta\bR \, .
\end{align}
In similar spirit to \eqref{stressref}, we define
\begin{align}
	\bar{\Sigma}^{kl} = \bar{A}^{ijkl}\epsilon_{ij} \, ,
\end{align}
and write the equations of equilibrium 
\begin{align}
	-\nabla_k\left( \sqrt{\bar{g}/g}\,\bar{\Sigma}^{kl} \nabla_l\bR \right) = {\bf{0}} \, , \label{bulkhybrid}
\end{align}
and free boundary conditions on $\partial V$ 
\begin{align}
 	\sqrt{\bar{g}/g}\,n_k \bar{\Sigma}^{kl} \nabla_l\bR  = {\bf{0}} \, . \label{bchybrid}
\end{align}
Projection of the divergence in equation \eqref{bulkhybrid} onto $\nabla^j\bR$ is also clearly a divergence, as in Section \ref{elasticitypresent}, namely
\begin{align}
	-\nabla_k\left(\sqrt{\bar{g}/g}\,\bar{\Sigma}^{kj}\right) = 0 \, , \label{bulkhybridproj}
\end{align}
which comes with boundary conditions
\begin{align}
 	\sqrt{\bar{g}/g}\,n_k \bar{\Sigma}^{kj}  = 0 \, . \label{bchybridproj}
\end{align}
I leave it as a tedious exercise to the reader to use identities such as $\sqrt{g}\,\Gamma^i_{ij}=\partial_j\sqrt{g}$ to show that, with somewhat loose interpretations of indices, the clean divergence \eqref{bulkhybridproj} is just the 
 non-divergence \eqref{bulkrefproj2} multiplied by a factor of $\sqrt{\bar{g}/g}$.  That is, the same functions are represented by these two expressions.  This relationship can be inferred from the treatment in Green and Zerna's foundational work from 1954 \cite{GreenZerna92}.  Concurrent and subsequent developments in continuum mechanics led to standard definitions of stress tensors and their interrelationships.  A summary of these and their connection with the present approach can be found in Appendix \ref{continuum}, which is probably best to read now.

\subsection{Remarks on geometric and material energies}\label{notgeometry}

A few additional remarks are in order on the distinction between geometric energies measured per unit volume and material energies measured per unit mass or reference volume.
Further related points will be made in Section \ref{furthernotgeometry}.

A commonly encountered geometric energy is the ``soap film'' energy or surface energy per unit present area (two-dimensional volume).
Consider a surface $\bX$ with normal $\uvc{N}$, and the energy
\begin{align}
	E_{\mathrm{sf}}  =   \int dA  =   \int \sqrt{a}\, dx^1dx^2 \, . \label{soapfilm}
\end{align}
For the variation we will need \cite{DesernoNotesDG} 
\begin{align}
	\delta\sqrt{a} &= \frac{1}{2}\frac{\delta a}{\sqrt{a}} = \frac{1}{2}\frac{1}{\sqrt{a}}\frac{\partial a}{\partial a_{\alpha\beta}} \delta a_{\alpha\beta} = \frac{1}{2}\frac{a\, a^{\alpha\beta}}{\sqrt{a}}\delta a_{\alpha\beta} =\sqrt{a}\, \nabla^\alpha\bX\cdot\nabla_\alpha\delta\bX \, ,\label{varareaform}
\end{align}
so that
\begin{align}
	\delta E_{\mathrm{sf}}  &=   \int \delta\sqrt{a}\, dx^1dx^2 \, , \nonumber \\
	&=   \int dA\, \nabla^\alpha\bX\cdot\nabla_\alpha\delta\bX \, , \nonumber \\
	&= - \int dA\, \nabla^2\bX \cdot \delta\bX \, \nonumber \\
	&\quad\,\, +  \oint dL\, n_\alpha \nabla^\alpha\bX \cdot \delta\bX \, ,
\end{align}
where $dL$ is a boundary element.  
The boundary term represents a tangential force resisting extension of the surface, while the bulk term is the Laplacian of the position vector $\nabla^2\bX = 2H\uvc{N}$,  
which seeks, for example, to collapse a sphere to a point.
Note that a similar energy for a three-dimensional body would have resulted in the vanishing bulk term $\nabla^2\bR={\bf{0}}$. 

While this type of energy is appropriate for describing a soap film, whose energy is contained in its soap-air interfaces, it is not suitable for describing kinetic or elastic energies.
In mathematical models of lipid membranes, the surface area is often held fixed with a constraint, while a 
 geometric energy is used to obtain simple and clean expressions \cite{Powers10, Deserno15}.
In this constrained limit, the difference between geometric and material energies arising from the variation of the area form can be absorbed into a Lagrange multiplier. 
Consider the soap film again, and let $\gamma$ be a multiplier.  Clearly, 
\begin{align}
  	\sqrt{g}\,\delta\left[ \mathcal{E}_0 + \gamma\left( \sqrt{g} - \sqrt{\bar{g}} \right) \right] = \sqrt{g}\left[ \delta\mathcal{E}_0 + \gamma\nabla^i\bR\cdot\nabla_i\delta\bR \right] 
\end{align}
and 
\begin{align}
	 \delta\left( \sqrt{g}\left[ \mathcal{E}_0 + \gamma\left( \sqrt{g} - \sqrt{\bar{g}} \right) \right] \right) = \sqrt{g}\left[ \delta\mathcal{E}_0 + \left(\gamma+\mathcal{E}_0\right)\nabla^i\bR\cdot\nabla_i\delta\bR \right]  
\end{align}  
are equivalent under the redefinition $\gamma \rightarrow \gamma + \mathcal{E}_0$.  
For elastic sheets, Guven and M{\"{u}}ller  \cite{GuvenMuller08} introduced something akin to the more restrictive rigid body constraint  $T^{ij}\left(\nabla_i\bR\cdot\nabla_j\bR - \bar{g}_{ij}\right)$, with multipliers $T^{ij}$ for all components of the metric.  Noting that $\nabla^i\bR\cdot\nabla_i\delta\bR = g^{ij}\nabla_i\bR\cdot\nabla_j\delta\bR$, the appropriate redefinition in this case is 
\begin{align}
	T^{ij} \rightarrow T^{ij} + \mathcal{E}_0g^{ij} \, . \label{absorb}
\end{align}
The reader may find it illustrative to examine equations (122-123) of \cite{Powers10} in this light.
In Section \ref{plates} we will see how, in the absence of such constraints on stretching, elastic bending energies differ in important ways from geometric bending energies.

\section{reduction}\label{reduction}

In this section, I will reduce the three-dimensional energy corresponding to the Almansi or present metric approach of Section \ref{elasticitypresent} to a two-dimensional energy defined on the mid-surface of a plate.
I will also consider the Biot strain in addition to the strains of Section \ref{elasticity}, and use
the simple case of an extensible \emph{elastica} or unidirectionally bent plate to demonstrate 
that the choice of strain measure has significant qualitative effects on the definition of bending energies.

The reduction process is not the focus of this paper, so no attempt is made at rigor.  It is included here for completeness of the plate equation derivations, but also in order to raise some important issues that are 
 likely to be far less 
  dependent on the specific reduction process than they are on the choice of strain measure.  
 Thus I will follow standard, if somewhat unjustified, procedures invoking the first and second Kirchhoff-Love assumptions \cite{Efrati09jmps},
 despite the fact that these are contradictory.
For a different method, see Steigmann \cite{Steigmann07, Steigmann08}.

\subsection{A reduced energy for a plate}\label{reductionplate}

This derivation is based on that of \cite{Dias11}, which follows that of \cite{Efrati09jmps}.
Let $\bX\left( \{x^\alpha\} \right)$ be a two-dimensional mid-surface of a three-dimensional plate $\bR\left(\{x^i\}\right)$.
Recall the convention, earlier implicitly introduced, for Latin indices $i \in \{1, 2, 3\}$ and Greek indices $\alpha \in \{1, 2\}$.  The latter coordinates are a subset of the former.  The Latin letter $z$ will denote the specific coordinate transverse to the plate, and should not be confused with a free index.
We write 
 $\bR\left(\{x^i\}\right) = \bR\left( \{x^\alpha\}, z \right)$ and $\bX\left( \{x^\alpha\} \right) = \bR\left( \{x^\alpha\}, 0 \right)$.
The goal is to express the energy as an effective field theory 
\begin{align}
	E(\bX) = \int dV \mathcal{E}(\bR) =  \int dA \int^{t/2}_{-t/2} dz\, \mathcal{E}(\bX, z) = \int dA\, \mathcal{E}_{2D}(\bX) \, , \label{energyrewrite}
\end{align}
where $dA = \sqrt{a} \prod\limits_{\alpha} dx^\alpha$, and $t$ is the uniform thickness of the undeformed plate.  
For compactness, we suppress explicit dependence on the reference configuration or metric in these expressions.

Presume a reference metric for the plate of the form
\begin{align}
	\bar{g}_{ij} = \begin{pmatrix} \left(\bar{g}_{\alpha\beta}\right)_{2\times2}&\bigcirc \\ \bigcirc&1\\ \end{pmatrix} = \begin{pmatrix} \left(\bar{a}_{\alpha\beta}\right)_{2\times2}&\bigcirc \\ \bigcirc&1\\ \end{pmatrix} \, . \label{referencemetric}
\end{align}
The second equality simply expresses the fact that $\bar{g}_{\alpha\beta}$ is not a function of $z$ for a plate, although it would be for a generic shell.
We may, but need not, think of this as coming from a rest configuration for a simple undeformed plate, $\bar{\bR}\left(\{x^i\}\right) = \bar{\bX}\left( \{x^\alpha\} \right) + z\bm{\mathrm{\hat{\bar{N}}}}$, where the reference unit normal $\bm{\mathrm{\hat{\bar{N}}}}$ is a constant vector.  
If this assumption is made, we can choose a very simple reference metric such as the two-dimensional Cartesian $\bar{a}_{11}=\bar{a}_{22}=1$, $\bar{a}_{12}=\bar{a}_{21}=0$.  
In the general case of an incompatible plate \cite{Efrati09jmps}, the three-dimensional plate does not have a reference embedding, but the mid-surface is still most likely embeddable in $\mathbb{E}^3$.

The first Kirchhoff-Love assumption diagonalizes the present metric,
\begin{align}
	g_{ij} = \partial_i \bR \cdot \partial_j \bR =   \begin{pmatrix} \left(g_{\alpha\beta}\right)_{2\times2}&\bigcirc \\ \bigcirc&g_{zz}\\ \end{pmatrix} \, . \label{presentmetricfirstKL}
\end{align}
We might think of this as arising from an assumed form for the position vector
$\bR\left(\{x^i\}\right) = \bX\left( \{x^\alpha\} \right) + z\partial_z\bR\left(\{x^i\}\right)$, where $\partial_z\bR$ is normal to $\bX$ and has a magnitude independent of $\{x^\alpha\}$.
In order to integrate over $z$, we need to rewrite the energy density in terms of $\bX$ and $z$ by getting rid of the third tensor index.
We can use the boundary condition \eqref{bcproj} on the top and bottom faces of the plate to obtain $\Sigma^{zj} = 0$ for any $j$.  Recall that the stress was defined in \eqref{stresspresent} as $\Sigma^{km} = A^{ijkl}\epsilon_{ij} \bar{g}^m_l$ with $A^{ijkl} = \lambda g^{ij}g^{kl} + \mu\left(g^{ik}g^{jl} + g^{il}g^{jk}\right)$.  Using $\Sigma^{zz}=0$ and the presumed form \eqref{referencemetric}, we find
$\lambda\epsilon^i_i + 2\mu\epsilon^z_z = 0$, and therefore
\begin{align}
	\epsilon^z_z = - \frac{\lambda}{\lambda+2\mu}\epsilon^\alpha_\alpha \, .\label{zzstrain}
\end{align}
The recasting of $\sqrt{\bar{g}/g}$ is complicated,\footnote{Those interested in working out the details will need to know that $\mathrm{det}\left(g_{\alpha\beta}\right) = \left(1 - 2zH+z^2K \right)\sqrt{a}\,$, where by $\mathrm{det}$ we mean the naive determinant of the components written as a matrix.} 
but for the moment we need only know that $\sqrt{\bar{g}/g} = \sqrt{\bar{a}/a} \left(1+O(\epsilon)\right)$.
Using the presumed forms \eqref{referencemetric} and \eqref{presentmetricfirstKL}, we can say that $\epsilon^i_j\epsilon^j_i = \epsilon^\alpha_\beta\epsilon^\beta_\alpha + \left(\epsilon^z_z\right)^2$ and rewrite the elastic energy \eqref{energydensitypresent},
\begin{align}
	\mathcal{E}\left(\bX,z\right) = \tfrac{1}{2}\sqrt{\bar{a}/a}\, \mu\left(\frac{2\lambda}{\lambda+2\mu}g^{\alpha\beta}g^{\gamma\zeta} + g^{\alpha\gamma}g^{\beta\zeta} + g^{\alpha\zeta}g^{\beta\gamma}\right)\epsilon_{\alpha\beta}\epsilon_{\gamma\zeta} + O\left(\epsilon^3\right) \, .
\end{align}

Now we need to express off-mid-surface terms in terms of mid-surface terms.
To simplify the calculations, we invoke the second Kirchhoff-Love assumption, which comes from a (clearly incorrect) presumed deformed configuration
$\bR\left(\{x^i\}\right) = \bX\left( \{x^\alpha\} \right) + z\uvc{N}\left(\{x^\alpha\}\right)$, such that
\begin{align}
	g_{ij} = \partial_i \bR \cdot \partial_j \bR =   \begin{pmatrix} \left(g_{\alpha\beta}\right)_{2\times2}&\bigcirc \\ \bigcirc&1\\ \end{pmatrix} \, , \label{presentmetricsecondKL}
\end{align}
where 
\begin{align}
	g_{\alpha\beta} = a_{\alpha\beta} - 2zb_{\alpha\beta} + z^2c_{\alpha\beta} \label{2dmetricin3d}\, .
\end{align}
We introduce $\varepsilon_{\alpha\beta}\left( \{x^\alpha\} \right) \equiv \epsilon_{\alpha\beta}\left( \{x^\alpha\}, 0 \right)$, so that $2\varepsilon_{\alpha\beta} = a_{\alpha\beta}-\bar{a}_{\alpha\beta}$ while $2\epsilon_{\alpha\beta} = g_{\alpha\beta}-\bar{g}_{\alpha\beta} = g_{\alpha\beta}-\bar{a}_{\alpha\beta} = 2\varepsilon_{\alpha\beta} - 2zb_{\alpha\beta} + z^2c_{\alpha\beta}$.  Note that surface indices are raised with the inverse surface metric, so while the bulk mixed component $\epsilon^\alpha_\beta = g^{\alpha\gamma}\epsilon_{\gamma\beta}$, the surface mixed component $\varepsilon^\alpha_\beta = a^{\alpha\gamma}b_{\gamma\beta}$, and similarly for $b^\alpha_\beta$ and $c^\alpha_\beta$.

Following Fl{\"{u}}gge \cite{Flugge72}, we approximate the inverse metric by first defining $\lambda^\beta_\alpha \equiv \delta^\beta_\alpha + zb^\beta_\alpha + z^2c^\beta_\alpha$ so that $g^{\alpha\beta} = \lambda^\alpha_\gamma\lambda^\beta_\zeta a^{\gamma\zeta} + O\left((zb)^3\right)$
and $g^{\alpha\beta}g_{\beta\gamma}=\delta^\alpha_\gamma + O\left((zb)^3\right)$.  
The mixed strains are
\begin{align}
	\epsilon^\alpha_\beta &= g^{\alpha\gamma}\epsilon_{\gamma\beta} \, , \nonumber \\
	&= \left( \delta^\alpha_\zeta + zb^\alpha_\zeta + z^2c^\alpha_\zeta \right)\left( \delta^\gamma_\eta +zb^\gamma_\eta + z^2c^\gamma_\eta \right)a^{\zeta\eta}\left( \varepsilon_{\gamma\beta} -zb_{\gamma\beta} +\tfrac{z^2}{2}c_{\gamma\beta} \right) + O\left((zb)^3\varepsilon\right) \label{mixedstrains} \, , \\
	&= \varepsilon^\alpha_\beta -zb^\alpha_\beta 
	+ O\left((zb)^2, zb\varepsilon\right) \, .
\end{align}
Following traditional practice, we will initially keep terms of orders $(zb)^2$, $zb\varepsilon$, and $\varepsilon^2$ in our quadratic strain energy, although terms odd in $z$ will integrate to zero.  However, we will revisit this choice in Section \ref{reductionextraterms}.
We write
\begin{align}
	\mathcal{E}\left(\bX,z\right) = \tfrac{1}{2}\sqrt{\bar{a}/a}\, \mathcal{A}^{\alpha\beta\gamma\zeta} \left(\varepsilon_{\alpha\beta} -zb_{\alpha\beta}\right)
\left(\varepsilon_{\gamma\zeta} -zb_{\gamma\zeta}\right)
 + O\left((zb)^3, (zb)^2\varepsilon, zb\varepsilon^2, \varepsilon^3\right) \, , \label{energydensitypresent3d}
\end{align}
where $\mathcal{A}^{\alpha\beta\gamma\zeta}= \mu\left(\frac{2\lambda}{\lambda+2\mu}a^{\alpha\beta}a^{\gamma\zeta} + a^{\alpha\gamma}a^{\beta\zeta} + a^{\alpha\zeta}a^{\beta\gamma}\right)$, 
and we are now forming invariants of two-dimensional surface tensors.
This approximate plate energy displays a clean separation between mid-surface stretching and bending energies,
\begin{align}
	\mathcal{E}_{2D}(\bX) &=  \int^{t/2}_{-t/2} dz\, \mathcal{E}(\bX, z) \, \nonumber \\
	&= \tfrac{1}{2}\sqrt{\bar{a}/a}\, \mathcal{A}^{\alpha\beta\gamma\zeta}\left( t\, \varepsilon_{\alpha\beta}\varepsilon_{\gamma\zeta} +\tfrac{t^3}{12}\,b_{\alpha\beta}b_{\gamma\zeta} \right) + O\left(t(tb)^3, t(tb)^2\varepsilon, t(tb)\varepsilon^2, t\varepsilon^3\right) \, ,  \label{energydensitypresent2d}
\end{align}
although the energies themselves are not entirely separate because of the compatibility constraints \eqref{Codazzi} and \eqref{Gauss}
relating metric and curvature tensors.
The bending contribution to the energy, arising from asymmetric off-mid-surface stretching, is
\begin{align}
	 \sqrt{\bar{a}/a}\,\tfrac{1}{2}\tfrac{t^3}{12}\mathcal{A}^{\alpha\beta\gamma\zeta}b_{\alpha\beta}b_{\gamma\zeta} &= \sqrt{\bar{a}/a}\,\tfrac{1}{2}\tfrac{t^3}{12}\,2\mu\left( \tfrac{2\left(\lambda+\mu\right)}{\lambda+2\mu}\,4H^2 - 2K\right) \, , \nonumber \\
	 &\equiv \sqrt{\bar{a}/a}\,\left( \BH H^2 - \BK K \right) \, ,
\end{align}
where we define the moduli $\BH\equiv\tfrac{4\left(\lambda+\mu\right)}{\lambda+2\mu}\,\BK$ and $\BK\equiv\tfrac{t^3\mu}{6}$ 
for compactness of subsequent presentation.  
These bending terms are distinct from those in Willmore \cite{Willmore65, PinkallSterling87}
 or Helfrich \cite{Helfrich73, Evans74, Jenkins77jmb} energies because of the prefactor $\sqrt{\bar{a}/a}\,$ that produces an elastic energy per unit mass.
The consequences of this distinction, including a contribution to the Euler-Lagrange equations from the elastic Gaussian term, will be shown in 
Section \ref{plates}.

\subsubsection{Extra terms in the energy}\label{reductionextraterms}

Actually, things need not be as simple as the typical derivation above implies.  Let us note an oddity.  
The nicely decoupled form \eqref{energydensitypresent2d} 
of the energy is the result of an approximation of the energy rather than the equations of equilibrium.
It will be seen in Section \ref{plates} that the variation of the bending terms, which are quadratic in curvature, will lead to terms cubic in curvature in the equations of equilibrium.
These are traditionally retained in the shape equation of Helfrich lipid membranes just as they are for \emph{elastica}, where indeed they need to be in order to include non-inflectional (double-well) shapes among the solutions.
Meanwhile, dropped terms of order $(zb)^2\varepsilon$ in \eqref{energydensitypresent3d} would have yielded comparable terms through the variation of the strain.\footnote{The importance of terms in the equations will also depend on derivatives.}
The same would be true for an energy constructed using the reference or hybrid approaches, and 
something similar can be said for the difference between geometric and material energies, as will be noted later in Section \ref{furthernotgeometry}.\footnote{Note that there should be no complications coming from the approximation of the volume prefactor $\sqrt{\bar{g}/g}\,$ as the area prefactor $\sqrt{\bar{a}/a}\,$, as long as we are adopting the second Kirchhoff-Love assumption. 
 Instead of dropping terms as higher-order in strain, as we did earlier, we may simply recognize that the term $\sqrt{\bar{g}}$  
is uniform in $z$ for a plate.}
 Recognizing this, Dias and co-workers \cite{Dias11} retained mixed stretching-bending terms in their derivation and showed that they could be absorbed, along with other higher-order terms arising from variation of the volume form in their geometric energy, into the definition of the stress.  
   I will show the derivation of an augmented stress that includes ``extra terms'' in Section \ref{platestretching}, which can be employed later in equations of equilibrium in Section \ref{platecombining}, but will neglect these terms as being of higher order when discussing energies in the following section.
   For a plate energy with even higher-order curvature and mixed terms, see \cite{DelfaniShodja16}.
   
Overall, something seems fishy about the expansion, and a possible reason will become apparent shortly in Section \ref{biotstrains}.

\subsection{Different bending energies for extensible \emph{elastica}}\label{extensibleelastica}

At this point, a simple example will illustrate the differences in the approaches of Sections \ref{elasticityreference}-\ref{elasticityhybrid} when used in conjunction with the Kirchhoff-Love assumptions. For simplicity of exposition, moduli, factors of two,
or similar inessentials will be neglected in the descriptions of energies.

Consider an extensible variant of the classical planar \emph{elastica}, endowed with a one-dimensional bending energy on a curve $\bX(x^1)$ with a single material coordinate.  This can be thought of as the mid-line of a two-dimensional object, or of a three-dimensional rod confined in the plane, or as a representation of the mid-surface of a unidirectionally bent plate in which the second body coordinate is ignored.  The only difference in these descriptions would appear in the bending modulus.  The bending energy will arise from mean curvature, as the Gaussian curvature is zero.
If $x^1$ is the arc length of $\bX$ in the reference configuration, then $\bar{a}_{11} = \left(\bar{a}^{-1}\right)^{11} = \sqrt{\bar{a}} = 1$ and $\bar{\Gamma}^1_{11} = 0$. 
  In the present configuration, let the tangents be expressed in Cartesians as $\partial_1\bX = \Lambda \begin{pmatrix} \cos\theta \\ \sin\theta \end{pmatrix}$, where $\Lambda$ is the stretch and $\theta$ is the tangential 
   angle in the plane.  Thus, $a_{11}=\Lambda^2$, $a^{11}=1/\Lambda^2$, $\sqrt{a}=\Lambda$, and $\Gamma^1_{11} = \partial_1\Lambda/\Lambda$.

Regardless of which metric is used to form invariants and determine an energy, the present--- not reference--- covariant derivative is what appears from the expansion of the metric \eqref{2dmetricin3d}.
This derivative isolates the bending (normal) term from the stretching (tangential) term when taking two derivatives of $\bX$.  That is, while $\partial_1\partial_1\bX = \partial_1\nabla_1\bX = \partial_1\Lambda \begin{pmatrix} \cos\theta \\ \sin\theta \end{pmatrix} + \Lambda\partial_1\theta \begin{pmatrix} -\sin\theta \\ \cos\theta \end{pmatrix}$, the covariant derivative $\nabla_1\partial_1\bX = \nabla_1\nabla_1\bX = \partial_1\partial_1\bX - \Gamma^1_{11}\partial_1\bX = \Lambda\partial_1\theta \begin{pmatrix} -\sin\theta \\ \cos\theta \end{pmatrix}$.  
The term $\nabla_1\nabla_1\bX\cdot\nabla_1\nabla_1\bX = b_{11}b_{11}$ will be acted on by either reference or present inverse metrics depending on the approach. 

The reference metric (Green strain) and hybrid approaches of Sections \ref{elasticityreference} and \ref{elasticityhybrid}, respectively, have the same bending contribution to the energy, taking the form
\begin{align}
	\int \sqrt{\bar{a}}\, dx^1\, \left(\bar{a}^{-1}\right)^{11}\nabla_1\nabla_1\bX\cdot\left(\bar{a}^{-1}\right)^{11}\nabla_1\nabla_1\bX = \int \sqrt{\bar{a}}\, dx^1 \left(\Lambda\partial_1\theta\right)^2 \, . \label{examplereference}
\end{align}
Consider a ring of elastic plate material, that is, a piece of material that wants to be linear but has been bent into a circle and had its ends attached, rather than a circular shell with rest curvature.
 For purely topological reasons, the two-dimensional ring cannot achieve its rest configuration through its thickness, but its mid-line can achieve its one-dimensional rest configuration by, for example, choosing a circle of the right radius.
Now examine the reference bending energy \eqref{examplereference}. Increasing the radius of the circle will make this energy \emph{increase}, as $\partial_1\theta$ will remain the same, but $\Lambda$ will increase.  One would need to conclude that given the same amount of material in a circular configuration, larger-radius circles have greater ``bending energy''.  A ring formed by inextensible bending of a plate would try to relax its bending energy by contracting to a smaller radius at the expense of some compression energy.  This is quite a strange definition of bending energy.  
Something akin to this issue was observed in the context of shells by Pezzulla and co-workers \cite{Pezzulla17}, who then counteracted its effects by employing a prefactor related to through-thickness strains, but this prefactor does not exist for plates and thus does not resolve the problem.

Instead, consider the present metric (Almansi strain) approach of Section \ref{elasticitypresent}, which applies the Laplacian $\nabla^2 = \nabla^1\nabla_1 = a^{11}\nabla_1\nabla_1$ to $\bX$ to obtain
\begin{align}
	\nabla^2\bX =\frac{\partial_1\theta}{\Lambda}\begin{pmatrix} -\sin\theta \\ \cos\theta \end{pmatrix} \, .
\end{align}
Note that $\nabla^2\bX = \tfrac{1}{\sqrt{a}}\partial_1\left(\sqrt{a}a^{11}\partial_1\bX\right) =\tfrac{1}{\Lambda}\partial_1\left(\tfrac{1}{\Lambda}\partial_1\bX\right) = \partial_s^2\bX$, where $\partial_s$ is the derivative with respect to the present arc length, such that $\partial_1=\Lambda\partial_s$ (the first step uses a form of the covariant Laplacian valid only when acting on a ``surface scalar''). 
The magnitude of this expression is $\nabla^2\bX\cdot\nabla^2\bX = \left(\partial_s\theta\right)^2$, where $\partial_s\theta$ is the Frenet curvature.  This is truly a geometric object, but the integral is still over the material coordinate, making the energy itself material rather than geometric.  The bending contribution to the energy is
\begin{align}
	\int \sqrt{a}\, \sqrt{\bar{a}/a}\, dx^1\, \nabla^2\bX\cdot\nabla^2\bX = \int \sqrt{\bar{a}}\, dx^1 \left(\frac{\partial_1\theta}{\Lambda}\right)^2 \, . \label{examplepresent}
\end{align}
Now the stretch appears in the denominator rather than the numerator, and an increase in the radius of a circle leads to a smaller present bending energy \eqref{examplepresent}.
A ring formed by inextensible bending of a plate would try to relax its bending energy by expanding to a larger radius at the expense of some stretching energy.  

Let us also consider another operation on plate material, namely taking an open arc of a circle and extending it to make a longer arc of the same radius.
This will lead to no change in \eqref{examplepresent}, and is thus defined as a pure stretch in the present bending energy case.
This conclusion might seem intuitively reasonable at first.  However, we will revisit this issue shortly below in Section \ref{biotstrains} with another strain measure, using again the two operations of expansion of a circle and extension of a circular arc.

Using the present metric, the stretching energy term $\left(a^{11}\left(a_{11}-\bar{a}_{11}\right)\right)^2 = \left(\frac{\Lambda^2-1}{\Lambda^2}\right)^2$ will saturate at very large stretches but increase without bound in compression, while using the reference metric, the corresponding stretching energy term  $\left(\left(\bar{a}^{-1}\right)^{11}\left(a_{11}-\bar{a}_{11}\right)\right)^2 = \left(\Lambda^2-1\right)^2$ goes up quadratically with stretch 
 but is bounded in compression.  Neither model is really intended to be used beyond small to moderate strains.

\subsubsection{Other strain measures, particularly Biot}\label{biotstrains}

An important note is in order on treatments of extensible rods and sheets in the mechanics and physics literatures.
A direct or Cosserat approach to such structures \cite{EricksenTruesdell57, GreenLaws66, GreenNaghdi74, Green74-1} views them as low-dimensional bodies with phenomenological stretching, bending, and twisting energies or constitutive relations, without consideration of the higher-dimensional origins of such terms.  Instead, simple kinematical variables and constitutive laws are postulated.
Neither of the quadratic Green or Almansi bulk energies considered thus far corresponds to the simplest direct theory of rods.

Antman \cite{Antman68-2}, Reissner \cite{Reissner72}, and Whitman and DeSilva \cite{WhitmanDeSilva74} formulated equilibrium equations for rods using generalized strain variables, one of which is $\partial_1\theta$.  
Huddleston \cite{Huddleston78} used 
similar variables,  
while Green and Laws \cite{GreenLaws66} used $\Lambda\partial_1\theta$, 
and other contemporaries developed different measures and approaches \cite{Tadjbakhsh66, Kafadar72, EpsteinMurray76compstruct}.
  Adopting Reissner's and Whitman and DeSilva's simple assumption of a linear relation between moment and $\partial_1\theta$ preserves the structure of the equations of equilibrium of the \emph{elastica}, 
 with the inextensible arc length derivative replaced by an extensible material derivative. 
Antman, without proposing any particular constitutive relation, favors the kinematical variable $\partial_1\theta$ because it is in some sense a measure of curvature that does not include changes arising from dilation (see page 98 of \cite{Antman05}),  
 such as expansion of a circle, which he sees as a ``pure extension'' of an initially curved rod  \cite{Antman68-2}.  
  
 A Reissner-like model will not arise if one performs a reduction of either Green or Almansi energies as in the preceding Section \ref{extensibleelastica}.
   An important, but apparently little known and frequently re-discovered, fact is that this model can arise from a reduction if the  
 Biot strain $\Lambda -1$, rather than the Green 
  strain $\tfrac{1}{2}\left(\Lambda^2-1\right)$, is used  \cite{IwakumaKuranishi84, Chaisomphob86, Magnusson01, IrschikGerstmayr09}.
The Biot strain is equivalent to the Green strain for small strains, $\Lambda \approx 1$, and corresponds to bending terms of the form $\left(\partial_1\theta\right)^2$.   
Such a bending energy does not change when a circle is expanded to change its radius, in keeping with Antman's preferred definition of a pure stretch, but increases when a circular arc is extended at fixed radius.
  Oshri and Diamant \cite{OshriDiamant17} also prefer this form for similar reasons as the early rod mechanicians, because it ``decouples'' the bending terms, so that one defines a moment linearly dependent on the kinematic bending variable and independent of stretch. 
While they consider pure bending, and recognize that the Green energy \cite{Efrati09jmps} produces a bending energy dependent on stretch, they don't notice that this dependence is counterintuitive, and do not consider extension of a bent arc.
Irschik and Gerstmayr \cite{IrschikGerstmayr09} explicitly show the nonlinear relationship between moment and stretch and bending variables corresponding to a  reduction from Green strains.
  
One can also consider strains corresponding to the use of the present basis, 
which would provide Almansi $\frac{1}{2}\frac{\Lambda^2-1}{\Lambda^2}$ and Swainger $\frac{\Lambda-1}{\Lambda}$ strains as counterparts to the Green and Biot strains, respectively. 
Unlike Green and Almansi, we do not currently have a covariant expression for Biot and Swainger energies in terms of derivatives for the general case.
Oshri and Diamant \cite{OshriDiamant17} have formulated a non-covariant two-dimensional theory for axisymmetrically deformed plates that employs Biot strains.
I will not attempt here to formulate a general covariant three-dimensional description in terms of Biot or Swainger strain in 3D, but one can infer that the Swainger bending energy will 
be qualitatively similar to the Almansi description, in which extending an arc with fixed radius is a pure stretch, whereas increasing a circle's radius involves both stretching and bending.  
To illustrate this line of thinking, consider constructing a scalar energy in one dimension from a variable such as $\left(\sqrt{a}-\sqrt{\bar{a}}\right)$, a tensor density equivalent to a component of the Biot strain.  
Curiously, the first work of which I am aware that demonstrates the Biot-Reissner correspondence also expresses the stretch in these terms \cite{IwakumaKuranishi84}, a trick that unfortunately does not generalize to higher dimensions.
  We can use other tensor densities to construct an invariant energy.  In particular, consider the terms
   $\left(\frac{1}{\sqrt{a}}\left(\sqrt{a}-\sqrt{\bar{a}}\right)\right)^2 = \left(\frac{\Lambda-1}{\Lambda}\right)^2$ and $\left(\frac{1}{\sqrt{\bar{a}}}\left(\sqrt{a}-\sqrt{\bar{a}}\right)\right)^2 = \left(\Lambda-1\right)^2$ constructed using the present and reference integration measures, respectively.
     These Swainger (present) and Biot (reference) energy terms have the same bounded or unbounded character in tension and compression as the analogous quartic-in-stretch Almansi and Green energies, but are quadratic in stretch.
A preference for blowup in compression rather than tension to describe real materials, something that happens naturally when using the present basis, led
 Magnusson and co-workers \cite{Magnusson01} to contrive a more complicated nonlinear constitutive relation than that corresponding to Biot energy.
 In another publication \cite{WoodHanna19}, my co-author and I confirm that a Swainger energy leads to the same $\left(\frac{\partial_1\theta}{\Lambda}\right)^2$ bending term as that obtained from the Almansi approach. 
 
Let us summarize here 
the behavior of bending energies, derived from quadratic bulk energies, associated with simple deformations of a plate.  
 Using the Green strain energy (\cite{Efrati09jmps} and its descendants), 
both expanding a circle and extending a circular arc increase the ``bending energy'', the former being a counterintuitive definition.  
Using the Biot strain energy (\cite{IrschikGerstmayr09} and others) leads to a Reissner-like model in which expanding a circle is a pure stretch that preserves ``bending energy'', while extending an arc increases it. 
 Using the Almansi or Swainger \cite{WoodHanna19} strain energies (Section \ref{elasticitypresent}), expanding a circle decreases the ``bending energy'', while extending an arc is a pure stretch that preserves it.  
Of course, all of these are constitutive assumptions so, intuitive or not, cannot be right or wrong except as descriptions of a particular material's experimentally determined behavior.

However, the simplicity of the Biot result suggests that a more appropriate expansion around the mid-surface is not in terms of Green strain or similar variables, but in something akin to Biot strain.
Some early workers 
 sought to compare effective field theories and direct approaches through expansions of the position vector \cite{AntmanWarner66, Green65, Green68, Green74-1, Green74-2}. 
 The use of a strain linear in displacement derivatives might even resolve the issue of inconsistency between approximating the energy and approximating the equations of equilibrium discussed in Section \ref{reductionextraterms}.
The choice of Green/Almansi components in models recently employed by the soft condensed matter community is likely driven both by the prominence of Green strain as the prototypical strain measure, as well as the ease in which such expansions can be constructed in terms of derivatives of position.
However, the use of linearized strain measures
based on linear springs and linear or nonlinear hinges in models of low-dimensional structures is also a popular approach in the same community.  
This conceptualization of a sheet as a continuum limit of a spring network, rather than a two-dimensional limit of a three-dimensional bulk elastic body, reflects the different origins of stretching and bending elasticity for molecular mesostructures and elastic solids and is likely the original driver behind Oshri and Diamant's use of Biot strains.
An influential paper by Seung and Nelson \cite{SeungNelson88} introduced a computational model for elastic sheets that has been used to explore the nature of crumpling singularities \cite{Witten07}, among other things.  They define a linearized stretching energy that is effectively Biot in nature. Their discretization of bending energy is nonlinear, and is intended to represent the continuum per area Helfrich energy, although the distinction between per mass, as naturally arises in the discretization, and per area, as treated in the theory, does not seem to be acknowledged.  This is perhaps consistent with the Landau-Lifshitzian neglect of in-plane nonlinearity \cite{LandauLifshitz86} in their theoretical discussion.
There is still disagreement on the correct bending energy for such discrete structures  \cite{SchmidtFraternali12, Kleiman16}, which sometimes employ linear hinges.
A theory with linear springs and linear hinges was applied to approximate extensible elastica by Oshri and Diamant \cite{OshriDiamant16}, who derive a bending term of the form $\left(\partial_1\theta\right)^2$, like that of the Reissner model.  
In a study of buckling of axisymmetric shells, 
Knoche and Kierfeld \cite{KnocheKierfeld11} use Biot-like strains and an equivalent measure for bending strains to obtain simple constitutive relations.

\subsubsection{Further remarks on geometric and material energies: Elasticity is not geometry}\label{furthernotgeometry}

While the 
 quantity $\frac{\partial_1\theta}{\Lambda}$ in \eqref{examplepresent} is the curvature $\partial_s\theta$, the integral is distinct from the geometric integral that one would obtain by integrating an energy per length (one-dimensional volume).
  For example, let the reference and present configurations of a body be two concentric circles parameterized such that the reference arc length and the tangential angle are always identical, $x^1=\theta$.  
Let the reference configuration be a circle of radius unity, such that the reference arc length coincides with the tangential angle $x^1=\theta$, so that $\bar{\bX}\left(\theta\right) = \begin{pmatrix} \cos\theta \\ \sin\theta \end{pmatrix}$ and the reference metric $\bar{a}_{\theta\theta} =  \partial_\theta\bar{\bX}\cdot\partial_\theta\bar{\bX}  = 1$.  
 Let the present configuration be a circle of radius $r$ given by $\bX\left(\theta\right) = r \begin{pmatrix} \cos\theta \\ \sin\theta \end{pmatrix}$ with present metric $a_{\theta\theta} = \partial_\theta\bX\cdot\partial_\theta\bX = r^2$.
The integral
\begin{align}
	\int_0^{2\pi}\sqrt{\bar{a}}\,d\theta\, H^2 &= \int_0^{2\pi} 1\,d\theta\, \frac{1}{r^2} = \frac{2\pi}{r^2} \, ,
 \end{align}
akin to \eqref{examplepresent}, is clearly different from the integral
\begin{align}
	\int_0^{2\pi}\sqrt{a}\,d\theta\, H^2 &= \int_0^{2\pi} r \,d\theta\, \frac{1}{r^2} = \frac{2\pi}{r} \, .
 \end{align} 
The significance for surfaces is even more qualitatively striking.  For a sphere, the geometric integral $\int H^2 dA$ is independent of radius, while  $\int H^2 d\bar{A}$ is not.  
Another geometric 
quadratic bending energy term would be $\int K dA\,$, which by the Gauss-Bonnet theorem is a combination of boundary terms and a topological invariant.  But the elastic energy $\int K d\bar{A}$ is not, with consequences that will be shown in the next section. 
Some of the elegant transformations, including exploitation of conformal invariance \cite{White73, Castro-VillarrealGuven07, GuvenVazquez-Montejo13}, 
 that are possible with geometric energies are not possible with elastic energies.

\section{plates}\label{plates}

In this section, I will perform the variation $\delta E = \delta \int dA\, \mathcal{E}_{2D}\,$, with the density $\mathcal{E}_{2D}$ defined in \eqref{energydensitypresent2d}, piece by piece, with steps shown explicitly and quantities written using derivatives of position.
While this particular form of the derivation is, to my knowledge, unique to this paper, 
the vector equations and boundary conditions are implicitly constructable from published results in component form \cite{CapovillaGuven02, Capovilla02, Guven04}, and other compact vector forms can be found in the literature \cite{HilgersPipkin92qjmam, HilgersPipkin92qam, HilgersPipkin96, Steigmann13ijnlm, Steigmann13ijnlmcorr}.  Aside from notational compactness, vector approaches naturally give rise to different boundary conditions 
than those obtained by breaking the vector variation into normal and tangential components, a point discussed elsewhere by Steigmann \cite{Steigmann13ijnlm}.

First we will require another basic tool.
As mentioned in Section \ref{elasticityreference}, the variation passes through the partial derivative, and thus through the first covariant derivative acting on $\bX$, but not subsequent covariant derivatives as are found in the curvature components.  Recall the variations of the metric and inverse metric, 
\begin{align}
	\delta a_{\alpha\beta} &= \nabla_\alpha\delta\bX\cdot\nabla_\beta\bX + \nabla_\alpha\bX\cdot\nabla_\beta\delta\bX \, , \label{varmetric} \\
	\delta a^{\alpha\beta} &= -\nabla^\alpha\bX\cdot\nabla^\beta\delta\bX - \nabla^\beta\bX\cdot\nabla^\alpha\delta\bX \, . \label{varinversemetric}
\end{align}
The variation of the curvature invariants or components uses the fact that $\delta\left(\nabla_\beta\nabla_\alpha\bX\right) = \delta\left(\partial_\beta\nabla_\alpha\bX - \Gamma^\gamma_{\beta\alpha}\nabla_\gamma\bX\right) = \partial_\beta\nabla_\alpha\delta\bX - \delta\left(\Gamma^\gamma_{\beta\alpha}\right)\nabla_\gamma\bX - \Gamma^\gamma_{\beta\alpha}\nabla_\gamma\delta\bX = \nabla_\beta\nabla_\alpha\delta\bX - \delta\left(\Gamma^\gamma_{\beta\alpha}\right)\nabla_\gamma\bX $.  This, along with $\delta\uvc{N}\cdot\uvc{N}=0$, leads to the curious result that
\begin{align}
	\delta b_{\alpha\beta} &= \delta\left(\nabla_\beta\nabla_\alpha\bX\right)\cdot\uvc{N} + \nabla_\beta\nabla_\alpha\bX\cdot\delta\uvc{N} \, , \nonumber \\
	&= \nabla_\beta\nabla_\alpha\delta\bX\cdot\uvc{N} \, , \label{curiousresult}
\end{align}
despite the variation \emph{not} passing through both covariant derivatives.  
This may be used 
 to write
\begin{align}
	\nabla^2\bX\cdot\delta\left(\nabla_\beta\nabla_\alpha\bX\right) = \nabla^2\bX\cdot\nabla_\beta\nabla_\alpha\delta\bX \, . \label{curiousresult2}
\end{align}

Note that in a recent work on lipid membrane bending elasticity, Capovilla \cite{Capovilla17} assumes a 
commutation (vanishing Lie bracket) between his variation and covariant derivative that allows for a compact general expression for Euler-Lagrange equations of energies depending on second covariant derivatives.  This is a restriction on the variation that is apparently allowable for fluid membranes, in which any in-plane strains can be compensated by a reparameterization.  Although results on variation of curvature invariants probably don't depend on this moot point, his equation (41) is not general and cannot be applied in the context of solid elastic membranes.

\subsection{Stretching or constraint}\label{platestretching}

The stretching term is $\tfrac{t}{2}\sqrt{\bar{a}/a}\, \mathcal{A}^{\alpha\beta\gamma\zeta} \varepsilon_{\alpha\beta}\varepsilon_{\gamma\zeta}$.
Following an identical procedure to that of Section \ref{elasticitypresent}, we can define stress 
 components
\begin{align}
	\varsigma^{\gamma\eta} &= t \mathcal{A}^{\alpha\beta\gamma\zeta} \varepsilon_{\alpha\beta} \bar{a}^\eta_\zeta \, , \\
	&= \mathcal{A}^{\alpha\beta\gamma\eta} \varepsilon_{\alpha\beta} + O(\epsilon^2) \, ,
\end{align}
such that the term in the bulk $A$ is
\begin{align}
	-\nabla_\gamma\left( \sqrt{\bar{a}/a}\, \varsigma^{\gamma\eta} \nabla_\eta\bX \right) \, , \label{bulk2Dstretch}
\end{align}
and the term on the boundary $\partial A$ is
\begin{align}
 	\sqrt{\bar{a}/a}\, n_\gamma \varsigma^{\gamma\eta} \nabla_\eta\bX  \, . \label{bc2Dstretch}
\end{align}
Terms of exactly the same form are obtained if we 
define $ \varsigma^{\gamma\eta}$ as a tensor multiplier constraining the metric of the mid-surface \cite{GuvenMuller08} and, in lieu of a stretching energy, add a constraint term of the form $\tfrac{1}{2}\varsigma^{\gamma\eta}\left(a_{\gamma\eta}-\bar{a}_{\gamma\eta}\right)$ to a bending energy $\mathcal{E}_0$.
This constraint restricts the deformations to isometric deformations of the mid-surface.
  In this limit, $\sqrt{a}=\sqrt{\bar{a}}$ and use of one or the other quantity will only redefine the multipliers in a manner akin to equation \eqref{absorb} in Section \ref{notgeometry}, 
 \begin{align}
	\varsigma^{\alpha\beta} \rightarrow \varsigma^{\alpha\beta} + \mathcal{E}_0a^{\alpha\beta} \, . \label{absorb2d}
\end{align}

\subsubsection{Extra terms in the stress}\label{plateextraterms}

Let us briefly see what would happen if we had retained extra terms, so as to approximate the equations of equilibrium rather than the energy, as discussed in Section \ref{reductionextraterms}.
Returning to the expression for the mixed strains \eqref{mixedstrains}, and using the symmetry of the second and third fundamental forms,
\begin{align}
	 \epsilon^\alpha_\beta &= \varepsilon^\alpha_\beta -zb^\alpha_\beta + 2zb^{\gamma\alpha}\varepsilon_{\gamma\beta} - \tfrac{3z^2}{2}c^\alpha_\beta + O\left((zb)^3, (zb)^2\varepsilon\right)
 \, .
\end{align}
The strain energy density is, using the symmetries of the elastic tensor,
\begin{align}
	\mathcal{E}\left(\bX,z\right) &= \tfrac{1}{2}\sqrt{\bar{a}/a}\, \mathcal{A}^{\alpha\beta\gamma\zeta} \left[ \left(\varepsilon_{\alpha\beta} -zb_{\alpha\beta}\right)
\left(\varepsilon_{\gamma\zeta} -zb_{\gamma\zeta}\right) -3z^2c_{\alpha\beta}\varepsilon_{\gamma\zeta} - 4z^2b_{\alpha\beta} b^\eta_\gamma \varepsilon_{\eta\zeta} \right] \nonumber \\
 &+ \mathrm{odd\,} (zb)^3\, \mathrm{terms} + O\left((zb)^4, (zb)^3\varepsilon, (zb)^2\varepsilon^2, \varepsilon^3\right) \, , \label{energydensitypresentaugmented3d}
\end{align}
where the odd terms will be integrated away.  We end up with
\begin{align}
	\mathcal{E}_{2D}(\bX) &=  \tfrac{1}{2}\sqrt{\bar{a}/a}\, \mathcal{A}^{\alpha\beta\gamma\zeta}\left[ \left( t\,\varepsilon_{\alpha\beta} - \tfrac{t^3}{4}c_{\alpha\beta}\right)\varepsilon_{\gamma\zeta} - \tfrac{t^3}{3}b_{\alpha\beta} b^\eta_\gamma \varepsilon_{\eta\zeta}	 +\tfrac{t^3}{12}\,b_{\alpha\beta}b_{\gamma\zeta} \right] \nonumber \\ &+ O\left(t(tb)^4, t(tb)^3\varepsilon, t(tb)^2\varepsilon^2, t\varepsilon^3\right) \, , \label{energydensitypresentaugmented2d}
\end{align}
which contains, alongside the familiar stretching and bending energies, two extra terms linear in the mid-surface strain.  Variation of the curvatures in these terms will lead to mixed higher-order terms, which we will neglect in the equations of equilibrium.  However, variation of the strain in these terms will lead to terms that combine with the variation of the stretching energy in an augmented stress
\begin{align}
	\tensor[^+]\varsigma{^{\gamma\eta}} &= \left[\mathcal{A}^{\alpha\beta\gamma\zeta}\left( \varepsilon_{\alpha\beta} - \tfrac{t^3}{8}c_{\alpha\beta} \right) + \mathcal{A}^{\alpha\beta\xi\zeta}\left(-\tfrac{t^3}{6}b_{\alpha\beta}b^\gamma_\xi \right) \right]\bar{a}^\eta_\zeta \, , \\
	&=  \mathcal{A}^{\alpha\beta\gamma\eta}\left( \varepsilon_{\alpha\beta} - \tfrac{t^3}{8}c_{\alpha\beta} \right) + \mathcal{A}^{\alpha\beta\xi\eta}\left(-\tfrac{t^3}{6}b_{\alpha\beta}b^\gamma_\xi \right) + O\left(t(tb)^2\varepsilon, \varepsilon^2\right) \, . \label{stresspresentaugmentedapprox}
\end{align}
The extra terms in this stress contribute terms to the equations of equilibrium similar to those arising from the variation of $H^2$.  
While it has been preferable to neglect smaller terms in our prior discussion of energies, their variational offspring are not really ``smaller'' than other terms we consider important.
We will return to this issue in Section \ref{platecombining} in an attempt to approximate the equations of equilibrium.
Note that while some of these terms would not arise when using the reference metric instead of the present metric, some such terms would indeed remain, 
 so these extra complications are relevant to either approach.
The difference between the expression \eqref{stresspresentaugmentedapprox} and its equivalent in \cite{Dias11} comes from not varying the volume form, or using its expansion off of the mid-surface, in the definition of a material energy.

\subsection{Mean curvature}\label{platemeancurvature}

The squared mean curvature term is $\sqrt{\bar{a}/a}\,\BH H^2$, with $H^2$ given by \eqref{meanX}.  
Recall that $\delta\left(dA\sqrt{\bar{a}/a}\right) = 0$.  We will use the trick \eqref{curiousresult2} and the symmetry of $\nabla_\alpha\nabla_\beta\bX$. 
Consider 
\begin{align}
	&\delta \int dA\, \sqrt{\bar{a}/a}\, \tfrac{1}{4}\nabla^2\bX\cdot\nabla^2\bX \nonumber \\
&= \int dA \sqrt{\bar{a}/a}\, \tfrac{1}{2} \nabla^2\bX \cdot \left[  a^{\alpha\beta}\delta\left(\nabla_\alpha\nabla_\beta\bX\right)+ \delta a^{\alpha\beta}\nabla_\alpha\nabla_\beta\bX \right] \, , \nonumber \\
	&= \int dA\, \sqrt{\bar{a}/a}\, \left[\tfrac{1}{2}\nabla^2\bX \cdot \nabla^2\delta\bX - \nabla^2\bX\cdot\nabla_\alpha\nabla_\beta\bX\nabla^\beta\bX\cdot\nabla^\alpha\delta\bX \right] \, , \nonumber \\
	&= \int dA \left[\nabla^2\left(\sqrt{\bar{a}/a}\,\tfrac{1}{2}\nabla^2\bX\right) + \nabla^\alpha\left(\sqrt{\bar{a}/a}\, \nabla^2\bX\cdot\nabla_\alpha\nabla_\beta\bX\nabla^\beta\bX \right) \right]\cdot\delta\bX \label{bulkmeancurvature} \\
	&\quad\,\, + \oint dL\, n_\alpha\left[ \sqrt{\bar{a}/a}\,\tfrac{1}{2}\nabla^2\bX \cdot\nabla^\alpha\delta\bX - \nabla^\alpha\left(\sqrt{\bar{a}/a}\,\tfrac{1}{2}\nabla^2\bX\right)\cdot\delta\bX \right. \nonumber \\
	&\quad\,\, \quad\,\, \quad\,\, \quad\,\, \quad\,\, \left. - \sqrt{\bar{a}/a}\,\nabla^2\bX\cdot\nabla^\alpha\nabla_\beta\bX\nabla^\beta\bX\cdot\delta\bX \right] \, , \nonumber \\
	&= \int dA \left[\nabla^2\left(\sqrt{\bar{a}/a}\,H\right)\uvc{N} - \sqrt{\bar{a}/a}\, H\nabla^2\uvc{N} \right]\cdot\delta\bX \nonumber \\
	&\quad\,\, - \oint dL\, n_\alpha \nabla^\alpha\left( \sqrt{\bar{a}/a}\, H \right) \uvc{N} \cdot\delta\bX \label{meanintact1} \\
	&\quad\,\, + \oint dL\, \sqrt{\bar{a}/a}\, H n_\alpha \nabla^\alpha\left(\uvc{N} \cdot \delta\bX\right) \, \label{meanintact2} \, , \\
	&= \int dA \left[\nabla^2\left(\sqrt{\bar{a}/a}\,H\right)\uvc{N} - \sqrt{\bar{a}/a}\, H\nabla^2\uvc{N} \right]\cdot\delta\bX \, , \label{bulkmeancurvature2} \\
	&\quad\,\, + \oint dL \left[ \sqrt{\bar{a}/a}\, H n_\alpha \nabla^\alpha\uvc{N} -n_\alpha \nabla^\alpha\left( \sqrt{\bar{a}/a}\, H \right) \uvc{N} \right] \cdot\delta\bX \label{meanbroken1} \\
	&\quad\,\, + \oint dL\, \sqrt{\bar{a}/a}\, H \uvc{N} \cdot n_\alpha \nabla^\alpha \delta\bX \, \, , \label{meanbroken2}
\end{align}
where $n_\alpha\nabla^\alpha$ is the projection of the covariant derivative normal to the boundary.
Smooth boundaries have been assumed, although no corner terms would arise from this energy anyway.
The two boundary terms correspond to forces
 and 
moments, the latter arising from the derivative normal to the boundary, and are in accordance with Steigmann's separation of variations \cite{Steigmann13ijnlm}.
Had we instead kept $\nabla^\alpha\left(\uvc{N} \cdot \delta\bX\right)$ intact on the boundary, as in 
lines (\ref{meanintact1}-\ref{meanintact2}), we would have obtained boundary terms like those \cite{Koiter66, Capovilla02, Deserno07, Dias11} 
that arise when variations are broken into tangential and normal components, or akin to geometrically linearized F{\"{o}}ppl-von K{\'{a}}rm{\'{a}}n approaches such as that in \cite{LandauLifshitz86}.
While these provide a simpler force boundary condition, they implicitly view $\nabla^\alpha\left(\uvc{N} \cdot \delta\bX\right)$ as somehow independent of $\delta \bX$. 

We can rewrite the bulk term \eqref{bulkmeancurvature2} 
using Gauss-Weingarten (\ref{GW1}-\ref{GW2}) and Codazzi \eqref{Codazzi} to evaluate the Laplacian of the normal, 
\begin{align}
	&\left[ \nabla^2\left(\sqrt{\bar{a}/a}\,H\right) + \sqrt{\bar{a}/a}\,2H\left(2H^2 - K\right) \right]\uvc{N} + \sqrt{\bar{a}/a}\,\nabla_\alpha\left(H^2\right)\nabla^\alpha\bX \, , \\
	&= \left[ \nabla^2\left(\sqrt{\bar{a}/a}\,H\right) + \sqrt{\bar{a}/a}\,2H\left(H^2 - K\right) \right]\uvc{N} + \sqrt{\bar{a}/a}\,\nabla_\alpha\left(H^2\nabla^\alpha\bX\right) \, ,
\end{align}
where the first line separates normal and tangential components, but the second line separates a Helfrich-like term from another term arising because of the lack of variation of the area form in the elastic integral.

\subsection{Gaussian curvature}\label{plategaussiancurvature}

The Gaussian curvature term is $-\sqrt{\bar{a}/a}\,\BK K$, with $K$ given by \eqref{GaussX}.
This is an elastic form of a geometric energy whose variation would be a pure divergence, by the Gauss-Bonnet theorem.
By contrast, the elastic energy contributes to the Euler-Lagrange equations. 

We will require the following funny identity that makes use of Gauss \eqref{KR}, Weingarten (\ref{GW1}-\ref{GW2}), and Codazzi \eqref{Codazzi}:
\begin{align}
	\nabla^2\nabla^\alpha\bX - \nabla^\alpha\nabla^2\bX &= \nabla^\beta\left(b^\alpha_\beta\uvc{N}\right) - \nabla^\alpha\left(b^\beta_\beta\uvc{N}\right) \, , \nonumber \\
	&= \left( \nabla^\beta b^\alpha_\beta - \nabla^\alpha b^\beta_\beta\right)\uvc{N} + \left( - b^\alpha_\beta b^{\beta\gamma} + b^\beta_\beta b^{\alpha\gamma} \right)\nabla_\gamma\bX \, , \nonumber \\
	&=R\indices{^\beta^\alpha_\beta_\gamma}\nabla^\gamma\bX \, , \nonumber \\
	&=K\nabla^\alpha\bX \, . \label{Knabla}
\end{align}
Along with this, 
we will use 
the symmetry of $\nabla_\alpha\nabla_\beta\bX$, 
the identity $\nabla_\alpha\bX\cdot\nabla_\beta\bX\nabla^\beta\bX = \nabla_\alpha\bX$, and the fact that $\nabla_\alpha\nabla_\beta\bX\cdot\nabla_\gamma\bX=0$.
Rather than evaluate the second term in $K$ similarly to the first, 
consider the entire expression \eqref{GaussX} together,
\begin{align}
	\delta &\int dA\, \sqrt{\bar{a}/a}\,  \left[ \tfrac{1}{2}\nabla^2\bX\cdot\nabla^2\bX - \tfrac{1}{2}\nabla^\alpha\nabla_\beta\bX\cdot\nabla_\alpha\nabla^\beta\bX \right] \nonumber \\
	&= \int dA\, \sqrt{\bar{a}/a}\, \left[ \nabla^2\bX \cdot \nabla^2\delta\bX -2\nabla^2\bX\cdot\nabla_\alpha\nabla_\beta\bX\nabla^\beta\bX\cdot\nabla^\alpha\delta\bX \right. \nonumber\\
	&\quad\,\, \quad\,\, \quad\,\, \quad\,\, \quad\,\, \quad\,\,  \left. - \nabla^\alpha\nabla_\beta\bX \cdot \nabla_\alpha\nabla^\beta\delta\bX + 2\nabla^\alpha\nabla_\beta\bX\cdot\nabla_\alpha\nabla_\gamma\bX \nabla^\gamma\bX\cdot\nabla^\beta\delta\bX \right] \, , \nonumber \\
	&= \int dA\, \sqrt{\bar{a}/a}\, \left[ \nabla^2\bX \cdot \nabla^2\delta\bX - \nabla^\alpha\nabla_\beta\bX \cdot \nabla_\alpha\nabla^\beta\delta\bX -2K\nabla_\alpha\bX\cdot\nabla^\alpha\delta\bX  \right] \, , \label{linestart} \\
	&= \int dA\, \sqrt{\bar{a}/a}\, \left[ \nabla_\alpha\left( \nabla^2\bX\cdot\nabla^\alpha\delta\bX - \nabla^\alpha\nabla_\beta\bX\cdot\nabla^\beta\delta\bX \right) - K\nabla_\alpha\bX\cdot\nabla^\alpha\delta\bX  \right] \, , \nonumber \\	
	&= \int dA\, \sqrt{\bar{a}/a}\, \left[ \nabla^2\left(\nabla^2\bX\cdot\delta\bX\right) - \nabla_\alpha\nabla^\beta\left( \nabla^\alpha\nabla_\beta\bX\cdot\delta\bX\right) +\nabla_\alpha\left(K\nabla^\alpha\bX\right)\cdot\delta\bX \right] \, . \label{linestart2}	
\end{align}
This is tantalizingly close to a divergence, but misses on two counts.  One is that the factor $\sqrt{\bar{a}/a}$ does not pass through the covariant derivative.  The other is that the $\delta\bX$ in the third term sits outside the divergence.  This is simply the result of not varying the volume form  (see \eqref{varareaform}).  The term $\delta\left(\sqrt{a}\,K\right) = \sqrt{a}\left(\delta K + K\nabla^\alpha\bX\cdot\nabla_\alpha\delta\bX\right)$, where $\delta K$ is just the bracketed quantity in \eqref{linestart2}, is indeed a divergence, and thus moveable to the boundary \cite{Capovilla02, Dias11, Deserno15}.
The first two terms of \eqref{linestart2} simplify as follows,
\begin{align}
	\nabla^2\left(\nabla^2\bX\cdot\delta\bX\right) - \nabla_\alpha\nabla^\beta\left( \nabla^\alpha\nabla_\beta\bX\cdot\delta\bX\right) &= \nabla_\alpha\nabla^\beta\left[ \left( \nabla^2\bX\delta^\alpha_\beta - \nabla^\alpha\nabla_\beta\bX \right) \cdot\delta\bX \right] \, , \nonumber \\
	&= \nabla_\alpha\nabla^\beta\left[ \left( 2H\delta^\alpha_\beta - b^\alpha_\beta \right) \uvc{N} \cdot\delta\bX \right] \, , \nonumber \\
	&= \left( 2H\delta^\alpha_\beta - b^\alpha_\beta \right) \nabla_\alpha\nabla^\beta\left( \uvc{N} \cdot\delta\bX \right) \, ,
\end{align}
where we have used Codazzi \eqref{Codazzi} to see that $\left( 2H\delta^\alpha_\beta - b^\alpha_\beta \right)$ is divergence-free.

There are several ways of manipulating the integrals.  We choose to return to the calculation and proceed thus,
\begin{align}
	&\int dA\, \sqrt{\bar{a}/a}\, \left[ \left( 2H\delta^\alpha_\beta - b^\alpha_\beta \right) \nabla_\alpha\nabla^\beta\left( \uvc{N} \cdot\delta\bX \right)  +\nabla_\alpha\left(K\nabla^\alpha\bX\right)\cdot\delta\bX \right] \nonumber \\
	&= \int dA \left[ \left( 2H\delta^\alpha_\beta - b^\alpha_\beta \right)\nabla^\beta\nabla_\alpha\left( \sqrt{\bar{a}/a}\, \right)\uvc{N} +\sqrt{\bar{a}/a}\,\nabla_\alpha\left(K\nabla^\alpha\bX\right) \right] \cdot\delta\bX \nonumber  \\
	&\quad\,\, + \oint dL \left( 2H\delta^\alpha_\beta - b^\alpha_\beta \right) \left[ \sqrt{\bar{a}/a}\,n_\alpha\nabla^\beta\left(\uvc{N} \cdot\delta\bX\right) - n^\beta\nabla_\alpha\left(\sqrt{\bar{a}/a}\,\right)\uvc{N} \cdot\delta\bX \right] \, , \nonumber \\
	&= \int dA \left[ \left( 2H\delta^\alpha_\beta - b^\alpha_\beta \right)\nabla^\beta\nabla_\alpha\left( \sqrt{\bar{a}/a}\, \right)\uvc{N} +\sqrt{\bar{a}/a}\,\nabla_\alpha\left(K\nabla^\alpha\bX\right) \right] \cdot\delta\bX \nonumber \\
	&\quad\,\, + \oint dL  \left[ l^\beta l_\gamma \nabla^\gamma\left( b^\alpha_\beta n_\alpha \sqrt{\bar{a}/a}\, \right)  - \left( 2H\delta^\alpha_\beta - b^\alpha_\beta \right) n^\beta\nabla_\alpha\left(\sqrt{\bar{a}/a}\,\right) \right] \uvc{N}\cdot\delta\bX \label{Gaussintact1} \\
	&\quad\,\, + \oint dL\,  \sqrt{\bar{a}/a}\,\left(2H-n_\alpha b^\alpha_\beta n^\beta\right)n_\gamma\nabla^\gamma\left(\uvc{N}\cdot\delta\bX\right)  \, , \label{Gaussintact2} \\
	&= \int dA \left[ \left( 2H\delta^\alpha_\beta - b^\alpha_\beta \right)\nabla^\beta\nabla_\alpha\left( \sqrt{\bar{a}/a}\, \right)\uvc{N} +\sqrt{\bar{a}/a}\,\nabla_\alpha\left(K\nabla^\alpha\bX\right) \right] \cdot\delta\bX \nonumber \\
	&\quad\,\, + \oint dL  \left[\sqrt{\bar{a}/a}\,\left(2H-n_\alpha b^\alpha_\beta n^\beta\right)n_\gamma\nabla^\gamma\uvc{N} - \left( 2H\delta^\alpha_\beta - b^\alpha_\beta \right) n^\beta\nabla_\alpha\left(\sqrt{\bar{a}/a}\,\right)\uvc{N} \right. \nonumber \\
	&\quad\,\, \quad\,\, \quad\,\, \quad\,\, \quad\,\, \left. + l^\beta l_\gamma \nabla^\gamma\left( b^\alpha_\beta n_\alpha \sqrt{\bar{a}/a}\, \right)\uvc{N}   \right] \cdot\delta\bX \label{Gaussbroken1} \\
	&\quad\,\, + \oint dL\,  \sqrt{\bar{a}/a}\,\left(2H-n_\alpha b^\alpha_\beta n^\beta\right)\uvc{N}\cdot n_\gamma\nabla^\gamma\delta\bX  \, . \label{Gaussbroken2}
\end{align}
The final lines use the decomposition
$\nabla^\beta = \left(n^\beta n_\gamma + l^\beta l_\gamma\right)\nabla^\gamma$, where $n_\gamma\nabla^\gamma$ and $l_\gamma\nabla^\gamma$ are the projections of the covariant derivative onto the unit normal to the boundary, and along the boundary, respectively, 
and $n_\alpha n^\alpha = 1$ and $n_\alpha l^\alpha = 0$.
 A quantity such as $l^\beta l_\gamma\nabla^\gamma T_\beta$ is a divergence on the one-dimensional boundary.    
Smooth boundaries are assumed, so  
a corner term $- l^\beta b^\alpha_\beta n_\alpha \sqrt{\bar{a}/a}\,\uvc{N} \cdot\delta\bX$ is thrown away.  Note that one of the terms in the corner jump condition of \cite{Dias11} is necessarily zero because $n_\alpha l^\alpha = 0$, but was kept to show kinship with the term in the moment balance equation.
Again, had we grouped terms as in the lines
 (\ref{Gaussintact1}-\ref{Gaussintact2}), we would have obtained boundary terms consistent with the form found in \cite{Koiter66, Capovilla02, Deserno07, Dias11, LandauLifshitz86}.
Other possibilities will be discussed in Section \ref{platecombining} just below.

\subsection{Combining and dropping terms}\label{platecombining}

The scale factor $\sqrt{\bar{a}/a}$ adds complications by behaving essentially like a variable bending modulus throughout these expressions.
We have retained it for a long time, primarily as a reminder that mass is conserved and that elasticity is per mass (material) and not per area (geometric). 
However, remembering that $\sqrt{\bar{a}/a} = 1 + O(\varepsilon)$, let's consider whether we can justify dropping the additional ugly terms.

Here I offer a hand waving argument that says we can do so, anticipating the final form of the equations of equilibrium, and focusing on the shape equation, the normal projection of the bulk terms.  If the scale factor were unity, we would end up with terms of orders $(t^3\nabla^2 b, t^3b^3, tb\varepsilon)$, the latter coming from the stretching term, while the augmentation of the stress adds nothing new.  If these terms are retained in the final balance, it implies that we consider $\varepsilon$, $tb$ and $t\nabla$ ``small''.  Inclusion of the scale factor gives us terms of the original orders multiplied by $\varepsilon$, as well as new terms of orders $(t^3 b \nabla^2\varepsilon, t^3\nabla b \nabla\varepsilon)$.  According to our thinking, these are also smaller by an order of $\varepsilon$, and can be dropped.

 After this, the only remaining terms that distinguish between per mass and per area energies are terms 
$\BH\nabla_\alpha\left(H^2\nabla^\alpha\bX\right)$ and $-\BK\nabla_\alpha\left(K\nabla^\alpha\bX\right)$ in the vector bulk equations.
These can be grouped with the stress tensor, although this makes one boundary condition particularly long-winded.
Note that the term involving $K$ will disappear in the limit of a mid-surface isometry only when the reference configuration is a flat plate; for incompatible elasticity 
 this term is nontrivial.
Had we approximated the scale factor in the energy, we could have gone further and constructed an approximate geometric energy, varied the area form $\sqrt{a}$ along with the rest of the bending energy, and obtained a pure boundary term for the variation of the Gaussian curvature, as is done in the isometric treatment of Guven and M{\"{u}}ller \cite{GuvenMuller08}.
As discussed in Sections \ref{reductionextraterms} and \ref{plateextraterms}, there is a difference between dropping terms in the energy and dropping them in the equations of equilibrium.  The difference between the geometric $\sqrt{a}\,K$ and the elastic $\sqrt{\bar{a}}\,K$ is another example of a $(zb)^2\varepsilon$ term that could have been added to or dropped from the energy, but contributes terms at retained orders in the equations of equilibrium.  Thus, particularly if we wish to retain the extra stress terms (see \eqref{stresspresentaugmentedapprox}), we may wish to keep this term in the equations of equilibrium while dropping all other terms involving the scale factor.
And even if we truncate the energy at quadratic order in mid-surface strain and curvature, it is most physical to write the energy as being per mass, which means retaining the bulk Gaussian term.  Of course, if the energy is written up to higher orders, these distinctions become important, and any approximation as a per area energy becomes less accurate.

In the limit of a mid-surface isometry, the full expression for the variation of $K$ simplifies considerably.  It is then a matter of taste whether one uses the area form $\sqrt{a}$ or the equivalent reference area form $\sqrt{\bar{a}}$ as the measure of integration, the difference appearing only in the multipliers.  The former choice, while unconventional from the point of view of the mechanics literature, is preferred by many physicists and leads to geometric integrals with elegant properties
\cite{GuvenMuller08}. 
However, this hides the material nature of the elastic energy, and does not cleanly link up with the material form required when mid-surface strains, particularly of higher orders, are included.
Thus, despite some loss of elegance, we favor retaining the per mass form of the equations even in the isometric limit.

To derive the equations for the isometric limit, $\tensor[^+]\varsigma{^{\alpha\beta}}$ must be replaced with a multiplier. 
Terms with a strain, and bending terms that arose from variation of a strain, no longer have meaning. 
 The relevant energy density is then
\begin{align}
	\int \sqrt{\bar{a}}\,dx^1dx^2 \left[ \left( \BH H^2 - \BK K \right) + \tfrac{1}{2}\tilde{\varsigma}^{\alpha\beta}\left(\nabla_\alpha\bX\cdot\nabla_\beta\bX - \bar{a}_{\alpha\beta}\right) \right] \, , \label{energypresentconstrained}
\end{align}
where $\tilde{\varsigma}^{\alpha\beta}$ is a tensor multiplier enforcing the constraint of isometric deformations of the mid-surface.
This is an elastic (non-geometric) version of the Guven-M{\"{u}}ller functional \cite{GuvenMuller08} that describes a metrically constrained Willmore-Helfrich energy.
Because $\sqrt{\bar{a}} = \sqrt{a}$ in the limit, the only difference between the resulting equations and those in \cite{GuvenMuller08} is a redefinition of the multipliers 
as in \eqref{absorb2d}.

Let us now write the combined equations using the symbol $\tau^{\alpha\beta}$, which should be interpreted either as the Lagrange multiplier $\tilde{\varsigma}^{\alpha\beta}$  in the constrained theory \eqref{energypresentconstrained}, or as the stress $\tensor[^+]\varsigma{^{\alpha\beta}}$ from \eqref{stresspresentaugmentedapprox} in an elastic theory that retains terms up to $O((t\nabla)^2(tb), (tb)^3, (tb)\varepsilon)$ in the normal projection of the bulk equations.

The combined bulk equations are 
\begin{align}
	\BH\left[ \nabla^2 H + 2H\left(H^2 - K\right) \right]\uvc{N} - \nabla_\alpha\left( \left[ \tau^{\alpha\beta} - \left(\BH H^2 - \BK K \right)a^{\alpha\beta} \right] \nabla_\beta\bX \right) = {\bf{0}} \, , \label{bulkcombined}
\end{align}
with free boundary conditions for forces
\begin{align}
	\left[-\BH n_\alpha\nabla^\alpha H - \BK l^\beta l_\gamma\nabla^\gamma\left( b^\alpha_\beta n_\alpha \right) \right]\uvc{N} +\left[\BH H  -\BK\left(2H-n_\alpha b^\alpha_\beta n^\beta\right)\right]n_\gamma\nabla^\gamma\uvc{N} + n_\alpha\tau^{\alpha\beta}\nabla_\beta\bX = {\bf{0}} \, , \label{forcebccombined}
\end{align}
and moments
\begin{align}
	\left[ \BH H - \BK\left( 2H - n_\alpha b^\alpha_\beta n^\beta \right) \right]\uvc{N} = {\bf{0}} \, .  \label{momentbccombined}
\end{align}
The bulk equation \eqref{bulkcombined} and force boundary condition \eqref{forcebccombined} correspond to the variation $\delta\bX$, and the moment boundary condition corresponds to the derivative of this variation normal to the boundary, $n_\alpha\nabla^\alpha\delta\bX$. 
It seems clear that these can be taken as independent variations, in keeping with the grouping of terms in lines (\ref{meanbroken1}-\ref{meanbroken2}) and (\ref{Gaussbroken1}-\ref{Gaussbroken2}).
Again, this is in contrast with those treatments \cite{Koiter66, Capovilla02, Deserno07, Dias11} that break the variation into tangential and normal components, and consider $\delta\bX$ and $n_\alpha\nabla^\alpha\left(\uvc{N}\cdot\delta\bX\right)$ to be independent variations, in keeping with the alternate grouping of terms in lines (\ref{meanintact1}-\ref{meanintact2}) and (\ref{Gaussintact1}-\ref{Gaussintact2}).
The present form is consistent with the approach of Steigmann \cite{Steigmann13ijnlm}, and contains additional $n_\gamma\nabla^\gamma\uvc{N}$ terms in the force boundary condition \eqref{forcebccombined} that are absent in
\cite{Koiter66, Capovilla02, Deserno07, Dias11, LandauLifshitz86}.
Note that the moment condition \eqref{momentbccombined} can be used to replace the terms multiplying $n_\gamma\nabla^\gamma\uvc{N}$ in \eqref{forcebccombined}, which simply vanish for a moment-free boundary.
We also did not group $-\left(\BH H^2 - \BK K \right)a^{\alpha\beta}$ terms with the stress in \eqref{forcebccombined} as was done in the bulk equation, because this merely complicates the force boundary condition--- compare the tangential projection of \eqref{forcebccombined} with equation (6) of \cite{Dias11}, where additional terms appear outside the stress.
For an elastic plate, $\BH$ is linearly related to $\BK$, so one can combine the mean curvature terms in the moment bracket, but we refrain from doing so in order to keep the results applicable to any general choice of coefficients for the two quadratic curvature energies.

The vector bulk equations \eqref{bulkcombined} are a surface divergence, which can be seen by looking back to line \eqref{bulkmeancurvature},
\begin{align}
	\nabla_\alpha \left[ \BH\left(\nabla^\alpha H\uvc{N} - H\nabla^\alpha\uvc{N}\right) - \tau^{\alpha\beta}\nabla_\beta\bX - \BK K\nabla^\alpha\bX \right] = {\bf{0}} \, .\label{bulkdivergence}
\end{align}
Using Gauss \eqref{Gauss} and some manipulations, we can write
\begin{align}
	n_\alpha K\nabla^\alpha\bX = -\left( 2H - n_\alpha b^\alpha_\beta n^\beta \right)n_\gamma\nabla^\gamma\uvc{N} - l^\beta l_\gamma \nabla^\gamma\left( b^\alpha_\beta n_\alpha \right)\uvc{N} + l^\beta l_\gamma \nabla^\gamma \left( b^\alpha_\beta n_\alpha \uvc{N} \right) \, , \label{rewritingboundaryterms}
\end{align}
so that ($-n_\alpha$ times) the quantity inside the divergence in \eqref{bulkdivergence} differs from the force boundary terms in
\eqref{forcebccombined} only by a boundary divergence.

The normal and tangential projections of the bulk equations \eqref{bulkcombined} are
\begin{align}
	\BH\left[ \nabla^2 H + 2H\left(H^2 - K\right) \right] - \left[ \tau^{\alpha\beta} - \left(\BH H^2 - \BK K \right)a^{\alpha\beta} \right]b_{\beta\alpha} &= 0 \, ,  \label{bulkcombinednormalproj}\\
	-\nabla_\alpha \left[ \tau^{\alpha\gamma} - \left(\BH H^2 - \BK K \right)a^{\alpha\gamma} \right] &= 0 \, . \label{bulkcombinedtangentialproj}
\end{align}
The divergence form of \eqref{bulkdivergence} and its tangential projection \eqref{bulkcombinedtangentialproj} are clearly anticipated from the discussion in \cite{GuvenMuller08}. 
Note that terms $\left[\left(\BH H^2 - \BK K \right)a^{\alpha\beta}\right]b_{\beta\alpha} = 2H \left(\BH H^2 - \BK K \right)$ in \eqref{bulkcombinednormalproj} arising from the lack of variation of the area form
are similar to, but not quite the same as,
 the cubic terms in the Helfrich contribution on the left.
The normal and tangential projections of the force boundary condition \eqref{forcebccombined} are 
\begin{align}
	-\BH n_\alpha\nabla^\alpha H - \BK l^\beta l_\gamma\nabla^\gamma\left( b^\alpha_\beta n_\alpha \right) &= 0 \, , \label{forcebcnormalproj} \\
	n_\gamma\left( -\left[\BH H  -\BK\left(2H-n_\alpha b^\alpha_\beta n^\beta\right)\right]b^{\gamma\eta} + \tau^{\gamma\eta} \right) & = 0 \, , \label{forcebctangentialproj}
\end{align}
where again we do not group the $-\left(\BH H^2 - \BK K \right)a^{\alpha\beta}$ term with the stress $\tau^{\alpha\beta}$, to keep the expression \eqref{forcebctangentialproj} in a simpler form.
The moment boundary condition \eqref{momentbccombined} has only its normal projection,
\begin{align}
	\BH H - \BK\left( 2H - n_\alpha b^\alpha_\beta n^\beta \right) = 0 \, .  \label{momentbcnormalproj}
\end{align}
We can use \eqref{momentbcnormalproj} in \eqref{forcebctangentialproj} to find that 
$n_\gamma\tau^{\gamma\eta} = 0$ for a moment-free boundary.

There are other ways to write these equations and conditions, and the related forms for lipid membranes, and no doubt some are nicer than what is found here.
Notable examples include the use of force and moment 
\cite{Jenkins77siam, Steigmann12, Steigmann13ijnlm, Steigmann13ijnlmcorr} or  closely related 
stress and torque tensors 
\cite{CapovillaGuven02, Capovilla02, TuOu-Yang08, GuvenMuller08}, 
and the 
 normal and geodesic curvature and torsion of the boundary
\cite{TuOu-Yang08, Deserno15}.

Interpreting the equations as constrained elastic equations derived from \eqref{energypresentconstrained}, it is instructive to compare them with those derived in the arc length ``gauge'' for the inextensible \emph{elastica} using the functional $\tfrac{1}{2}\int ds \left[ \B\kappa^2 + \varsigma\left(\partial_s\bX\cdot\partial_s\bX - 1\right) \right]$, where $\kappa^2 = \partial^2_s\bX\cdot\partial^2_s\bX$ is the squared Frenet curvature.  Here $s$ is both the rest and present arc length, no metric tensor or covariant derivative explicitly appears, and $ds$ is not subject to variation.  Comparison with the variation of this functional shows us that $\tau^{ss} = \varsigma + 2\B\kappa^2 = \varsigma + 4\BH H^2$, with $\kappa = 2H$ and $2\B = \BH$.

\section{Additional discussion}\label{discussion}

The results derived here can just as well be thought of as applying to a plate with in-plane incompatibility, that is, an object with no through-thickness variation in rest metric.  An interpretation in terms of a reference embedding in $\mathbb{E}^3$ is no longer possible, although one could likely construct a smooth or piecewise smooth \cite{GemmerVenkataramani13} isometric embedding of the mid-surface.
The present metric (Almansi) approach, in concert with the second Kirchoff-Love assumption, leads to expressions in the Euler-Lagrange equations involving geometric quantities of the deformed surface.
Using another approach that favors the reference metric, it should be possible to derive similarly simple expressions involving the geometric features of the reference configuration.  This question would be better explored in the context of shells rather than plates.
The retention of the divergence form of the equations after tangential projection, a structure made clear in early work on bulk elasticity \cite{GreenZerna92} and in recent
approaches to solid surfaces \cite{GuvenMuller08}, 
 likely reflects a connection between the statements of conservation of momentum and pseudomomentum \cite{Singhdraft}.
 
The concept of an energy quadratic in strain has been ill defined.  If metrics or the $\epsilon_{ij}$ derived from them are seen as fundamental fields, then a general quadratic energy would contain terms constructed from both $g^{ij}\bar{g}_{jk}$ and $\bar{g}^{IJ}g_{jk}=\left(\bar{g}^{-1}\right)^{ij}g_{jk}$.  
Interestingly, one finds both $\bar{g}^{IJ}g_{ij}$ and $g^{ij}\bar{g}_{ij}\sqrt{g/\bar{g}}$ as commonly discussed invariants in early foundational work on elasticity \cite{GreenZerna92, DoyleEricksen56}, because one can rewrite invariants of right and left Cauchy-Green tensors and their inverses, and by extension the Green and Almansi strains, in terms of each other.  These translations involve the third (cubic) invariant, the Jacobian $\sqrt{g/\bar{g}}\,$.
But there are yet other measures of strain.  
The principal stretches are square roots of the eigenvalues of the right Cauchy-Green tensor, so something quadratic in the latter is quartic in the former.
Commonly used rubber elasticity models such as Neo-Hookean or Mooney-Rivlin energy densities explicitly contain terms quadratic in stretches, although they are not general expansions in stretch as might satisfy a physicist.

The difference in the importance or order of terms between the energy and equilibrium equations, leading to potential confusion as to when to drop terms, may simply indicate that an expansion in terms of mid-surface Green or Almansi strain and the product of curvature and thickness
is simply not the natural one.
Further indication of this possibility comes from the correspondence between Biot strain and the simplest direct theories of rods.
Thus, developing a general covariant theory of Biot strain in the language of physics would be valuable.
It would also connect with early treatments of low-dimensional bodies 
\cite{AntmanWarner66, Green65, Green68, Green74-1, Green74-2} and recent developments in soft matter \cite{KnocheKierfeld11, OshriDiamant16, OshriDiamant17}.
Many direct theories of rods and shells naively assume a simple relationship between a convenient kinematical variable and the stored energy function without determining what that possibly very complex function might be.
Some comparisons have been made between direct and reduced theories  \cite{AntmanWarner66, Green65, Green68, Green74-1, Green74-2, Parker79-2, OReilly98, Steigmann99mms}, but  it would be valuable to take a broader look at the simple generalized strains that naturally arise in Cosserat theories \cite{EricksenTruesdell57, Green65, GreenLaws66, GreenNaghdi74} 
and how they do, or do not, correspond to simple stored energy functions.
Linking a direct theory to a bulk elastic model not only justifies its employment, but could be a basis for derivation of better models for elastic strips or moderately thick structures.

\section{Conclusions}

I have compared and discussed some recent approaches to incompatible elasticity in the soft condensed matter physics community, and presented the derivation of plate equations in a compact form.
Among the issues raised were the meaning behind what physicists refer to as metric choices, the divergence form of equations of equilibrium, 
qualitative differences in derived bending energies and their predictions for a simple combined stretching and bending problem, the possible advantages of the Biot strain as an alternate measure to serve as a basis for expansions,
and the differences between geometric and material bending energies.

\appendix

\section{Relating to modern continuum mechanics}\label{continuum}

This appendix serves to indicate how to translate between the bulk-elastic quantities in the main text and those favored by modern continuum mechanics.  An example of a standard text in the latter format is \cite{GurtinFriedAnand10}.

For compactness and ease of comparison, let us introduce the deformation gradient $\bF$ and related two-point tensors, 
\begin{equation}
	\bF = \partial_i\bR\partial^I\bar{\bR} \quad,\quad \bF^T = \partial^I\bar{\bR}\partial_i\bR  \quad,\quad  \bF^{-1} = \partial_i\bar{\bR}\partial^i\bR  \quad,\quad \bF^{-T} = \partial^i\bR\partial_i\bar{\bR} \,  \quad,\quad
	\label{defgrad}
\end{equation}
whose purpose is essentially to transform the basis vectors in the two configurations into each other.  
Note that $\bar\nabla\bR = \bF$ while $\nabla\bR = \bI$, where the two nablas $\bar\nabla() = \bar\nabla_I()\partial^I\bar{\bR}$ and $\nabla() = \nabla_i()\partial^i\bR$ represent the gradient in the reference and present configurations, respectively.  

A fundamental object is the Cauchy traction $\bT(\uvc{n})$, the force per present area on the surface element $\uvc{n}dS$ of a material element in the present configuration.
The Cauchy stress $\bm\sigma$ is defined such that $\uvc{n}\cdot\bm\sigma= \bT$.  Transforming this surface element to its corresponding referential element by $\uvc{n}dS = \sqrt{g/\bar{g}}\,\bF^{-T}\cdot\bm{\mathrm{\hat{\bar{n}}}}d\bar{S}$ serves to define the first Piola-Kirchhoff stress, a two-point tensor such that $\bP\cdot\bm{\mathrm{\hat{\bar{n}}}}d\bar{S} = \uvc{n}dS\cdot\bm\sigma$.  Thus, 
\begin{align}
\bP = \sqrt{g/\bar{g}}\,\bm\sigma^T\cdot\bF^{-T} \quad \mathrm{and} \quad \bm\sigma = \sqrt{\bar{g}/g}\,\bF\cdot\bP^T \, .  \label{stressdefinitions}
\end{align}

Equations for elastostatics in the absence of body forces or boundary tractions may be derived from varying either representation of an elastic energy, 
\begin{align}
	\delta\bar{E} = \int\! d\bar{V}\, \delta\bar{\mathcal{E}} = \int\! dV \sqrt{\bar{g}/g}\, \delta\bar{\mathcal{E}} \, .\end{align}
In general, the variation of the energy density will take the form $\delta\bar{\mathcal{E}}={\bf f}^I\cdot\partial_i\delta\bR$.  
The upper index is here written with a capital letter, implying that the energy density $\bar{\mathcal{E}}$ was constructed using tensors in the reference configuration, but this need not be the case, and this index will be paired with a different basis in the reference and present balances. 
These come from
\begin{align}
	\delta\bar{E} &= \int\! d\bar{V}\, {\bf f}^I\cdot\bar\nabla_I\delta\bR &=& \int\! dV \sqrt{\bar{g}/g}\, {\bf f}^I\cdot\nabla_i\delta\bR \, , \\
	&= \int\!d\bar{S}\, \bar{n}_I {\bf f}^I\cdot\delta\bR - \int\! d\bar{V}\, \bar\nabla_I {\bf f}^I\cdot\delta\bR &=& \int\!dS\, n_i \sqrt{\bar{g}/g}\, {\bf f}^I\cdot\delta\bR -  \int\! dV\, \nabla_i\left( \sqrt{\bar{g}/g}\, {\bf f}^I \right)\cdot\delta\bR \, . 
\end{align}
Identifying the first Piola-Kirchhoff stress $\bP = {\bf f}^I\partial_i\bar{\bR}$ and the Cauchy stress $\bm\sigma = \sqrt{\bar{g}/g}\,\partial_i\bR{\bf f}^I$, the corresponding bulk terms are
\begin{align}
	\bar\nabla\cdot\bP^T \quad \mathrm{or} \quad \nabla\cdot\bm\sigma \, ,
\end{align}
with boundary terms
\begin{align}
	\bm{\mathrm{\hat{\bar{n}}}}\cdot\bP^T \quad \mathrm{or} \quad \uvc{n}\cdot\bm\sigma \, .
\end{align}
The transpose is being used so that the referential divergence acts on the referential ``leg'' of the two-point tensor. 
The two descriptions represent force densities per reference or per present volume and area.
They can be translated between using the definitions \eqref{stressdefinitions}, the transformations of surface elements, and the transformations of the divergences
\begin{align}
	\sqrt{g/\bar{g}}\,\nabla\cdot() = \bar\nabla\cdot\left(\sqrt{g/\bar{g}}\,\bF^{-1}\cdot()\right) \quad \mathrm{and} \quad \sqrt{\bar{g}/g}\,\bar\nabla\cdot() = \nabla\cdot\left(\sqrt{\bar{g}/g}\,\bF\cdot()\right) \, .
\end{align}

Sections \ref{elasticityreference} and \ref{elasticityhybrid} make use of a St.Venant-Kirchhoff energy density $\bar{\mathcal{E}} = \tfrac{1}{2}\bar{\Sigma}^{KL}\epsilon_{kl}$, where $\bar{\Sigma}^{KL} = \bar{A}^{IJKL}\epsilon_{ij}$.
For this energy, we can identify ${\bf f}^K = \bar{\Sigma}^{KL}\partial_l\bX$, $\bP^T = \bar{\Sigma}^{KL}\partial_k\bar{\bR}\partial_l\bR$, and $\bm\sigma=\sqrt{\bar{g}/g}\,\bar{\Sigma}^{KL}\partial_k\bR\partial_l\bR$.  We can also define the weighted Cauchy stress $\sqrt{g/\bar{g}}\,\bm\sigma$ and the second Piola-Kirchhoff stress $\bS =  \bar{\Sigma}^{KL}\partial_k\bar{\bR}\partial_l\bar{\bR}$, such that $\bP = \bF\cdot\bS$. 
 In terms of components, 
 \begin{align}
 	(\bP^T)^{Kl} = \sqrt{g/\bar{g}}\,\sigma^{kl} = S^{KL} = \bar{\Sigma}^{KL} \, .
\end{align}
 These tensors all have the same components but different tensorial legs.  They are related to each other by operations such as transposition and application of the deformation gradient $\bF$ and its companions.  
 Much of the machinery of modern continuum mechanics is involved with switching legs on tensors.  Additionally, many related expressions have been derived for stresses as derivatives of energies with respect to strains or the deformation gradient \cite{Sansour93}.

Another simple and common example in soft matter elasticity is the incompressible neo-Hookean energy density $\bar{\mathcal{E}} = \tfrac{1}{2} \mu \left(\bar{g}^{IJ}g_{ij}-3\right) - p\left(\sqrt{g/\bar{g}}\,-1\right)$, which penalizes $\mathrm{Tr}(\bC)$ subject to the isochoric constraint $\sqrt{g/\bar{g}}\,-1$ enforced by a multiplier $p$.
Here, using $\delta \sqrt{g/\bar{g}}\, = \sqrt{g/\bar{g}}\, \nabla^i\bR\cdot\partial_i\delta\bR$,  we find ${\bf{f}^I} = \left[\mu \bar{g}^{IJ} - pg^{ij}\right]\partial_j\bR$ and, using the constraint, $\sqrt{g/\bar{g}}\,\bm\sigma = \bm\sigma = \left[\mu\bar{g}^{IJ} - pg^{ij}\right]\partial_i\bR\partial_j\bR  = \mu\bB -p\bI $.

\section*{Acknowledgments}

Conversations with M A Dias, E Hohlfeld, and H G Wood have been particularly helpful.
I thank S S Antman and A Yavari for clarifying some aspects of their work for me.
I also thank the organizers of the Kavli Institute for Theoretical Physics (NSF PHY-1748958) program on Geometry, Elasticity, Fluctuations, and Order in 2D Soft Matter for providing me a brief window of time and space to think on some of these issues, and the organizers of the Variational Models of Soft Matter conference in Santiago de Chile for the opportunity to discuss them.

\bibliographystyle{unsrt}


\end{document}